\documentclass[12pt]{article}
\pdfoutput=1
\usepackage{jheppub_X}

\usepackage[T1]{fontenc}
\usepackage[utf8]{inputenc}
\usepackage{cleveref}
\crefname{table}{Table}{Tables}
\crefname{equation}{Eq.}{Eqs.}
\crefname{appendix}{App.}{Apps.}
\crefname{section}{Sec.}{Secs.}
\crefname{figure}{Fig.}{Figs.}
\usepackage{amsmath}
\usepackage{setspace}
\usepackage{graphicx}
\usepackage[export]{adjustbox}

\usepackage{bm}
\usepackage{xfrac}
\usepackage{pbox}
\usepackage{comment}
\usepackage{slashed}
\usepackage{enumitem}
\usepackage[compat=1.0.0]{tikz-feynman}

\usepackage[most]{tcolorbox}
\definecolor{light-gray}{gray}{0.9}
\usepackage{bbold}

\usepackage[normalem]{ulem}

\makeatletter
\g@addto@macro\bfseries{\boldmath}
\makeatother

\tikzset{vertexstyle/.style={circle,draw,thick,inner sep=2pt}}
\tikzset{dotstyle/.style={thick,dotted}}
\tikzset{propstyle/.style={ultra thick}}

\newcommand{\nn}{\nonumber}

\newcommand{\be}{\begin{equation}}
\newcommand{\ee}{\end{equation}}
\newcommand{\bea}{\begin{eqnarray}}
\newcommand{\eea}{\end{eqnarray}}

\newcommand{\ab}[1]{\langle #1 \rangle}
\newcommand{\sq}[1]{[ #1 ]}

\newcommand{\la}[1]{\langle #1 |}
\newcommand{\ls}[1]{\lbrack #1 |}
\newcommand{\ra}[1]{| #1 \rangle }
\newcommand{\rs}[1]{| #1 \rbrack}

\definecolor{colorTC}{rgb}{.2,.7,.2}

\newcommand{\s}{\hspace{0.8pt}}

\newcommand{\D}{\text{d}}

\preprint{CERN-TH-2024-069}

\title{\Large Recursion for Wilson-line Form Factors}

\author[a,b,c]{Timothy Cohen\s}
\author[a]{and Marc Riembau\s}

\affiliation[a]{\fontsize{10}{10}\selectfont Theoretical Physics Department, CERN, 1211 Geneva, Switzerland}
\affiliation[b]{\fontsize{10}{10}\selectfont Theoretical Particle Physics Laboratory, EPFL, 1015 Lausanne, Switzerland}
\affiliation[c]{\fontsize{10}{10}\selectfont Institute for Fundamental Science, University of Oregon, Eugene, Oregon 97403, USA}

\emailAdd{tim.cohen@cern.ch}
\emailAdd{marc.riembau@cern.ch}

\abstract{
Matrix elements of Wilson-line dressed operators play a central role in the factorization of soft and collinear modes in gauge theories.  When expressed using spinor helicity variables, these so-called form factors admit a classification starting from a Maximally Helicity Violating configuration, in close analogy with gauge theory amplitudes.  We show that a single-line complex momentum shift can be used to derive recursion relations that efficiently compute these helicity form factors at tree-level: a combination of lower point form factors and on-shell amplitudes serve as the input building blocks.  We obtain novel compact expressions for the $1\to 2$ and $1\to 3$ splitting functions in QCD, which also serves to validate our methods.
}

\begin{document}
\maketitle
\flushbottom
\setcounter{page}{2}

\begin{spacing}{1.1}
\parskip=0ex

\section{Introduction}
\label{sec:Introduction}
It is well known that an energetic charged particle does not exist in isolation.  It is instead surrounded by a cloud of soft and collinear radiation.  This observation plays a key role when exploring the factorization properties of QCD~\cite{Collins:1989gx}, and is critical to the success of the Soft Collinear Effective Theory (SCET)~\cite{Bauer:2001ct, Bauer:2000yr, Bauer:2001yt}, see~\cite{Becher:2014oda, Becher:2018gno, Cohen:2019wxr} for reviews.  This physical picture is captured by matrix elements of Wilson line dressed operators, which interpolate a hard charged particle.  These matrix elements are typically called ``form factors'' in the literature~\cite{Sen:1981sd,Collins:1989bt,Collins:1989gx,Brandhuber:2010ad, Brandhuber:2011tv, Feige:2013zla,Agarwal:2021ais}.  These objects have a wide variety of applications.  They serve as a key component of SCET calculations, for example appearing in the ``jet functions''~\cite{Bauer:2000yr} (after removing the contribution from soft radiation).  Form factors are also the input for QCD ``splitting functions''~\cite{Altarelli:1977zs, Campbell:1997hg, Badger:2004uk, Bern:2004cz, Badger:2015cxa, Czakon:2022fqi, Catani:2003vu, DelDuca:2019ggv, DelDuca:2020vst, Sborlini:2013jba, Sborlini:2014mpa, Sborlini:2014kla}, which are an important probe of the soft and collinear divergence structure of gauge theories and have applications for higher order calculations~\cite{Bern:1998sc, Kosower:1999rx, Bern:1999ry, Catani:2011st, Magnea:2018hab}. 

Our goal in this work is to explore the properties of the simplest class of form factors that appear in gauge theories.  This study will be facilitated by the introduction of a novel recursive approach to calculating form factors at tree level.  Recursive approaches to computing quantum field theory amplitudes have become established as a significantly more efficient methodology over Feynman diagrams, especially for the case of massless particles, see \cite{Dixon:1996wi, Elvang:2013cua, Dixon:2013uaa, Henn:2014yza, Cheung:2017pzi, Badger:2023eqz} for reviews.  One of the key organizing principles is to express the color ordered amplitudes in terms of spinor helicity variables, and to compute amplitudes for each choice of helicity for the massless external particles. In particular, this leads to a natural organization for QCD processes by first identifying the ``maximally helicity violating'' (MHV) amplitudes~\cite{Parke:1986gb, Mangano:1987xk}.  The MHV amplitudes have a particularly simple form, and for some recursive approaches serve as the input building blocks for the construction of the rest of the amplitudes, organized using a ``next-to-$\cdots$-next-to-MHV'' (N$^k$MHV) classification.  We show that the same approach reveals simple and universal properties of form factors.\footnote{For other work applying spinor helicity methods in the context of SCET, see~\cite{Moult:2015aoa, Kolodrubetz:2016uim, Feige:2017zci, Bhattacharya:2018vph}.}  We identify an MHV helicity form factor, which takes a remarkably simple form, inherited from the fact that such object arise as the collinear limit of on-shell amplitudes \cite{DelDuca:1999iql,Birthwright:2005ak,Birthwright:2005vi}.  The MHV helicity form factor (along with the MHV expansion for the amplitudes) serve as the starting point of a recursive methodology.

Recursion relations can be derived as a consequence of the Cauchy residue theorem. By deforming amplitudes into the complex plane, one can show that the physical amplitude of interest can be reconstructed by summing the factorizable contributions due to the presence of poles and a (potential) non-factorizable contribution that lives along the contour at infinity. To prove that a higher point object can be constructed from lower point inputs alone requires showing that the non-factorizable contribution vanishes.  There are a number of recursion relations for amplitudes, \emph{e.g.}~the BCFW~\cite{Britto:2004ap, Britto:2005fq}, CSW~\cite{Cachazo:2004kj}, and Berends-Giele~\cite{Berends:1987me} recursion relations.  The key difference is the choice of how to deform the physical amplitude into the complex plane; BCFW relies on introducing a complex shift for two of the external momenta, while CSW shifts all the external lines, see~\cite{Cohen:2010mi} for a study of all-line shifts for general theories.  In exactly the same spirit, we show that helicity form factors can be constructed recursively from lower point inputs.  One novelty of our form factor recursion relations is that they are derived as a consequence of a \emph{single-line} shift.

We demonstrate the power of our recursion relations for helicity form factors by explicitly computing a variety of non-trivial results.  Another well known aspect of recursion relations for amplitudes is that one can find very different expressions by implementing different shifts, which are nonetheless completely equivalent.  We will show that the same is true of the recursive approach introduced here.  As a validation of our methods, we provide new compact representations for some of the $1 \to 2$~\cite{Altarelli:1977zs} and $1 \to 3$~\cite{Catani:1998nv, Catani:1999ss} QCD splitting functions, providing a highly non-trivial check on our results.  Furthermore, in the recursive approach, these objects serve as the inputs for higher point splitting functions.

The rest of this paper is organized as follows.  In \cref{sec:Recursion}, we set up the form factor formalism, we introduce the one-line shift recursion relations, and we give a proof that higher point form factors can be fully constructed using lower point form factors and on-shell amplitudes as inputs.  Then we show how to apply the recursion relations to QED in \cref{sec:QED} and QCD in \cref{sec:QCD}, with an emphasis on the MHV classification.  We provide some outlook and future directions in \cref{sec:Conclusions}.  In \cref{sec:SplittingFuncs}, we show how to compute splitting functions from the helicity form factors.

\section{Recursion Relations for Helicity Form Factors}
\label{sec:Recursion}
In this section, we begin by reviewing the definition of the Wilson line form factors, which will allow us to introduce the matrix elements of interest in this paper.  We then review the spinor helicity formalism for light-cone coordinates, which provides us with a convenient set of variables with which to define the helicity form factors. With these building blocks in hand, we then introduce a class of complex momentum shifts that facilitates a recursive approach to computing the helicity form factors.  This is followed by an argument that the helicity form factors can be computed recursively from lower point building blocks.

\subsection{Wilson Line Preliminaries}

The form factors of interest here take the generic form
\be
\mathcal{F}^{\mathcal{O}}_\eta (\alpha)= \la{\alpha} W_\eta^\dagger(0) \mathcal{O}(0)\ra{0} \ ,
\label{eq:FormFactorDef}
\ee
where $\la{\alpha}$ is some multiparticle state with momentum $q^\mu$, $\mathcal{O}(x)$ is a local operator that transforms in a representation of the gauge group, and $W_\eta^\dagger(x)$ is a Wilson line in the direction $\eta^\mu$ whose transformation is fixed by the requirement that the form factor is gauge invariant:
\be
W_\eta^\dagger(x) = \text{P} \exp\bigg( ig\int_0^\infty \D t \,\eta\cdot A(x+t\eta) \bigg) \ ,
\ee
where P denotes path ordering, $g$ is a gauge coupling, $A^\mu(x)$ is a gauge field, and $t$ is an affine parameter.  Note that the momentum $q^\mu$ injected into the operator must be equal to the total momentum of the external state. 

We will consider operators $\mathcal{O}$ that interpolate single particle states: a scalar field $\phi$, a fermionic field $\Psi$, and a covariant derivative of a Wilson line $D_\mu W_\eta$. The corresponding form factors with single particle states, at leading order, are given by the free wavefunctions that are associated with the external massless state:
\begin{subequations}
\begin{align}
\la{\phi_q} W_\eta^\dagger  \phi^* \ra{0} &= 1\ , \label{eq:scalarWF}\\[5pt]
\la{f_q} W_\eta^\dagger  \overline{\Psi} \ra{0} &= \bar{u}(q)\ , \label{eq:fermionWF}\\[2pt]
\la{\gamma_q } \frac{i}{g}W_\eta^\dagger D^\mu W_\eta \ra{0} &= \epsilon^{\mu}(q)^* \label{photonWF} \ ,
\end{align}%
\label{eq:treeformfactors}%
\end{subequations}%
for the free scalar $\phi$, fermion $f$, and gauge boson $\gamma$ with momentum $q^\mu$.

Form factors of the type given in Eqs.~(\ref{eq:treeformfactors}) appear when formulating factorization theorems for soft and collinear modes in gauge theories \cite{Sen:1981sd,Collins:1989bt,Collins:1989gx,Feige:2013zla,Agarwal:2021ais}.
The physical setting is that some hard process produces a particle (the scalar, fermion or vector as determined by the choice of operator $\mathcal{O}(x)$) moving in the direction $\bar{\eta}^\mu$. 
This particle emits collinear radiation leading to a multiparticle final state $\la{\alpha}$ composed of collinear and soft quanta, whose total momentum is $q^\mu$. The direction $\bar{\eta}^\mu$ is given by $\bar{\eta}^\mu= (1,\vec{q}/|\vec{q}|)$.
The collinear radiation about this direction is determined by the form factor in \cref{eq:FormFactorDef}, which is gauge invariant due to the inclusion of the Wilson line.  Specifying that a gauge boson is radiated by a given particle is not a gauge invariant statement. Physically, the Wilson line encodes the radiation from the rest of the scattering process; the state traveling in the $\bar{\eta}$-collinear direction sees a charged classical source moving in an opposite direction given by $\eta^\mu= (1,-\vec{q}/|\vec{q}|)$.

One of our primary goals is to provide a novel helicity decomposition of the form factors that can be constructed recursively at tree-level.  The natural variables for expressing these objects are Weyl spinors.  This is the topic we discuss next.


\subsection{Light-cone Spinor Helicity}
Without loss of generality, we take the null direction $\eta^\mu$ to be along the $z$-axis.  
We call the parity-opposite null direction $\bar{\eta}$.
We also need to define the two transverse directions $\eta_\perp$ and $\eta_\perp^*$.  We take the explicit basis to be 
\begin{subequations}
	\begin{align}
	\eta^\mu&=(1,0,0,1)\,,
	&\eta^\mu_\perp = (0,-1,-i,0)\,,\\[3pt]
	\bar{\eta}^\mu&=(1,0,0,-1)\,,
	&\eta^{*\mu}_\perp = (0,-1,+i,0)\,.
	\end{align}
\end{subequations}
In terms of these basis elements, and using the mostly minus metric, the components of the momenta are given by
\begin{subequations}
	\begin{align}
	\bar{\eta}\cdot p &= p^0+p^3\equiv p^+\,,
	&\eta_\perp \cdot p = p^1+ip^2\equiv p_\perp\,, \\[3pt]
	\eta\cdot p &=p^0-p^3\equiv p^-\,,
	&\eta^*_\perp\cdot p = p^1-ip^2\equiv p_\perp^*\,.
	\end{align}
\end{subequations}
A generic lightlike momentum $p^\mu=E(1,s_\theta c_\phi,s_\theta s_\phi,c_\theta)$ becomes collinear to $\eta^\mu$ in the small $\theta$ angle limit.  In this limit, its components scale as 
\begin{align}
(p^+,p^-,\vec{p}_\perp)\sim p^+(1,\theta^2,\theta) + O(\theta^3) \, .
\end{align}
In other words, $\theta$ is the usual collinear expansion parameter in SCET.  The form factors studied here capture the collinear limit of amplitudes at leading order in $\theta$.

Since the vectors $\eta^\mu$, $\bar{\eta}^\mu$ are null, it is natural to associate them with two-component Weyl spinors, 
\begin{align}
\ra{\eta} = \sqrt{2}\begin{pmatrix}
1 \\ 0
\end{pmatrix}
\,, \hspace{20pt}
\ra{\bar{\eta}} = \sqrt{2}\begin{pmatrix}
0 \\ 1
\end{pmatrix}
\,, \hspace{20pt}
\rs{\eta} = \sqrt{2}\begin{pmatrix}
0 \\ 1
\end{pmatrix}
\,, \hspace{20pt}
\rs{\bar{\eta}} = \sqrt{2}\begin{pmatrix}
-1 \\ 0
\end{pmatrix}\,,
\label{eq:etaSpinors}
\end{align}
where we are using the spinor helicity conventions in~\cite{Schwartz:2014sze}.  The transverse directions are not null, and therefore it is not possible to associate a single Weyl spinor with them.  However, all the $\eta$-basis vectors do have a simple representation in terms of the spinors in Eqs.~(\ref{eq:etaSpinors}) when they are contracted with a gamma matrix:
\begin{subequations}
\bea
\slashed \eta &=& \ra{\eta}\ls{\eta} + \rs{\eta}\la{\eta}\,,\qquad\qquad
\slashed \eta_\perp = \ra{\eta}\ls{\bar{\eta}} + \rs{\bar{\eta}}\la{\eta}\,,\\[3pt]
\bar{\slashed \eta} &=& \ra{\bar{\eta}}\ls{\bar{\eta}} + \rs{\bar{\eta}}\la{\bar{\eta}}\,,\qquad\qquad
\slashed \eta^*_\perp = \ra{\bar{\eta}}\ls{\eta} + \rs{\eta}\la{\bar{\eta}}\,.
\eea%
\label{eq:etaslashed}%
\end{subequations}%
We can use these to write the collinear fermionic projector familiar from SCET as
\be
P_\eta = \frac{\slashed \eta \bar{\slashed \eta}}{4} = \frac{\ra{\eta}\la{\bar{\eta}}+\rs{\eta}\ls{\bar{\eta}}}{2} \, .
\label{eq:FermionProjectors}
\ee
When applied to an arbitrary spinor $\ra{p} = \sqrt{2E}(c_{\theta/2}, \,\, s_{\theta/2} e^{i\phi})^\text{T}$, the projector $P_\eta$ extracts the leading component of the spinor in the $\eta$ direction. For the basis defined above, $P_\eta \ra{p} = \sqrt{2E}c_{\theta/2} \ra{\eta} \simeq \sqrt{p^+} \ra{\eta}$.

We will also need the formalism for gauge bosons.
Note that the metric can be written as 
\be
2 g^{\mu\nu} =\eta^\mu\bar{\eta}^\nu + \bar{\eta}^\mu\eta^\nu-\eta^\mu_\perp\eta^{*\nu}_\perp-\eta^{*\mu}_\perp \eta^\nu_\perp\,.
\ee
The direction of the polarization associated with a vector propagating in the $\eta^\mu$ direction is, to leading order in $\theta$, perpendicular to both $\eta^\mu$ and $\bar{\eta}^\mu$, and perpendicular/parallel to $\eta^\mu_\perp$ and $\eta^{*\mu}_\perp$ depending on the helicity of the vector.
Therefore, the analog of the fermionic projectors in \cref{eq:FermionProjectors} for the vectors is given by
\be
P_{\eta^+}^{\mu\nu} = -\frac{\eta^\mu_\perp\eta^{*\nu}_\perp}{2}\,,
\qquad\text{and}\qquad
P_{\eta^-}^{\mu\nu} = -\frac{\eta^{*\mu}_\perp \eta^\nu_\perp}{2}\,.
\label{eq:vectorprojectors}
\ee
Indeed, one can check that $P_{\eta^+}^{\mu\nu}$ projects the leading component of a polarization vector corresponding to a plus helicity state 
\be
\epsilon^\mu_+(p) = \frac{1}{\sqrt{2}}\frac{\ls{p}\gamma^\mu\ra{\xi}}{\ab{\xi p}}\,,
\ee
($\ra{\xi}$ is a reference spinor)
as follows:
\be
P_{\eta^+}^{\mu\nu}\epsilon^\mu_+(p) \propto -\frac{\ls{p}\slashed \eta_\perp\ra{\xi}}{\ab{\xi p}} \eta^{*\nu}_\perp = -\frac{\sq{p\bar{\eta}}\ab{\eta \xi}}{\ab{\xi p}} \eta^{*\nu}_\perp \,=\, \eta^{*\nu}_\perp + O(\theta^2) \,,
\ee
where we used $\ra{p}= \sqrt{p^+}\ra{\eta} + O(\theta^2)$ and $\sq{p\bar{\eta}}\sim \sqrt{p^+}$.

\subsection{Helicity Form Factors}
We now use the notation introduced in the previous section to define the ``helicity form factors'' that are the central object of interest in this work, see \cite{Brandhuber:2010ad, Brandhuber:2011tv} for related studies.
If the operator is a Dirac fermion field $\Psi = (\ra{\Psi}, \rs{\Psi})$, we can define two scalar operators by projecting the left and right chiral parts of the field using $P_\eta$ defined in \cref{eq:FermionProjectors}: 
\begin{align}
\mathcal{O}^{f^-} = \langle \Psi(x) \ra{\eta}\,,
\qquad \text{and} \qquad 
\mathcal{O}^{f^+} = [\Psi(x)\rs{\eta}\,. 
\end{align}
The labels $f^-$ and $f^+$ indicate that the operator interpolates minus and plus helicity fermions respectively at leading order. 
Then, the form factors can be defined to be
\begin{subequations}
	\bea
	\mathcal{F}^{f^-}_\eta (\alpha)\,&=&\,\la{\alpha} W_\eta^\dagger\, \mathcal{O}^{f^-}\ra{0}\,,\\[3pt]
	\mathcal{F}^{f^+}_\eta (\alpha)\,&=&\,\la{\alpha} W_\eta^\dagger\,\mathcal{O}^{f^+}\ra{0}\,,
	\label{eq:fermionFFdef}
	\eea
\end{subequations}
where $\la{\alpha}$ is some multiparticle state. At leading order, $\mathcal{F}^{f^-}(f_1^-) = \ab{1\eta}$; the interpretation is that the form factor gives the leading component of the negative helicity fermion wavefunction in the $\bar{\eta}$ direction.
Also note that the projection eliminates $\mathcal{F}^{f^-}(f_1^+)$, and therefore the projected form factor only interpolates a negative helicity fermion.

The gluon form factors for each helicity are
\begin{subequations}
	\bea
	\mathcal{F}^{g^-}_\eta (\alpha)\,&=&\, \la{\alpha} \frac{i}{g} W_\eta^\dagger \frac{\eta_\perp\cdot D}{\sqrt{2}} W_\eta\ra{0}\,,\\[4pt]
	\mathcal{F}^{g^+}_\eta (\alpha)\,&=&\, \la{\alpha} \frac{i}{g} W_\eta^\dagger \frac{\eta^*_\perp\cdot D}{\sqrt{2}} W_\eta\ra{0}\,,
	\eea%
	\label{eq:gluonFFdef}%
\end{subequations}%
where $D^\mu$ is a gauge covariant derivative.
At leading order, the unprojected gluon form factor is proportional to the vector polarization with the arbitrary reference spinor replaced by the Wilson line direction, $\frac{\ra{p}\ls{\eta}}{\sq{\eta p}}$. We can instead choose some other direction for the Wilson line $\eta^\prime$, which can always be decomposed as $\rs{\eta^\prime} = a \rs{\eta} + b \rs{p}$, for some $a$ and $b$. If we express the form factor along the $\eta'$ direction, then the new form factor shifts by a vector proportional to $p^\mu$ as compared to the form factor defined in the original $\eta$ direction.  As long as the matrix element obeys the Ward identity, this extra term proportional to $\rs{p}$ does not contribute to physical observables, and the matrix element contracted with the form factor is independent of the Wilson line direction.

Therefore, given a $\bar{\eta}$-collinear photon or gluon, the leading order form factor for gauge bosons is given by
\be
\mathcal{F}^{g^+}_\eta (f_1^+)\,=\,\frac{\sq{1\eta}}{\ab{1\eta}}
\,,\qquad \text{and}\qquad
\mathcal{F}^{g^-}_\eta (f_1^-)\,=\,\frac{\ab{1\eta}}{\sq{1\eta}}\,.
\label{eq:gluonFFdef1point}
\ee
We emphasize that these are just the standard polarization vectors expressed in spinor helicity notation, where the arbitrary reference spinor is given by $\eta$.
So each projection leads to an operator that sources only one polarization.
All other components of the form factor are at least of order $\theta^1$ in the SCET power counting. 
Note that one consequence of defining the helicity form factors is that the little group scaling of the $\eta$ dependence in the resulting expressions corresponds to the helicity of the hard parton interpolated by the projected operator.


\subsection{Recursion Relations for Amplitudes}
\label{sec:recrelationAmps}
We begin this section by reviewing the recursive techniques for on-shell helicity scattering amplitudes, see \emph{e.g.}~\cite{Elvang:2015rqa}.  Our generalization to helicity form factors is then straightforward to explain.

The expansion of perturbative amplitudes in terms of Feynman diagrams is known to include redundancies that obscure the underlying analytic structure.
At tree level, the amplitude is a meromorphic function of the external kinematics and can therefore be completely determined from its poles.
It is possible to take advantage of this fact to design a recursive algorithm to compute the tree level amplitudes in an efficient, non-redundant way by constructing higher order amplitudes out of lower order building blocks.
This both provides new insight into the structure of the amplitudes, while also resulting in a multitude of new results that would not have been feasible to derive in terms of Feynman diagrams.

A recursion relation for tree-level scattering amplitudes can be obtained by deforming or ``shifting'' the external on-shell momenta $p^\mu_i$ into the complex plane, while maintaining the on-shell condition.
The shifted momentum is given by
\be
\hat{p}^\mu_i = p^\mu_i+z\s r^\mu_i\,,
\label{eq:pShift}
\ee 
where $z$ is a complex number, and $r^\mu_i$ are a set of vectors that define the shift.  The amplitude with shifted momenta $\hat{\mathcal{A}}$ then depends on $z$. 

When computing amplitudes, a class of recursion relations can be derived by enforcing that the $r^\mu_i$ obey the following properties:

\begin{center}
Momentum shift for amplitudes
\end{center}
\vspace{-20pt}
\begin{align}
(1) \quad \quad r_i\cdot r_j&=0 \qquad\text{for any pair, including } i = j\,;\nn \\[3pt]
(2) \quad \quad p_i\cdot r_i &= 0 \qquad \text{for all } i\,;\label{eq:amponshellconditions} \\[3pt]
(3) \quad \hspace{5pt} \sum_i r^\mu_i&=0\,.\nn
\end{align}
The reason for each condition is as follows.  Since we are considering tree-level amplitudes with massless particles, there will be physical poles $I$ when $P_I^2 = 0$ with $P^\mu_I=p^\mu_{i_1}+\dots p^\mu_{i_n}$, where the different set of momenta $\{i\}$ that appear accommodates all of the allowed channels of the process under consideration.  Then condition (1) implies that the poles of the shifted amplitudes are linear in $z$, such that $\hat{P}_I^2 = -P^2_I(z-z_I)/z_I$, with $z_I=-P_I^2/2P_I\cdot R_I$, and $R^\mu_I=r^\mu_{i_1}+\dots r^\mu_{i_n}$.
In other words, condition (1) implies that the shifted amplitude only includes single poles in the variable $z$.  Condition (2) ensures that the shifted momenta are on-shell, $\hat{p}_i^2=0$.
Finally, condition (3) is necessary in order to ensure total momentum conservation of the amplitude with the shifted momenta, $\sum_i\hat{p}^\mu_i=0$.

One then shifts the momenta that appear in the original tree-level amplitude $\mathcal{A}$ to define the shifted amplitude $\hat{\mathcal{A}}(z)$.
Using the Cauchy theorem, it is possible to write the $n$-point unshifted amplitude $\mathcal{A} \equiv \hat{\mathcal{A}}(z=0)$ in terms of shifted lower point amplitudes:
\be
\mathcal{A} = \sum_{z_I} \hat{\mathcal{A}}_L(z_I)\frac{1}{P^2_I}\hat{\mathcal{A}}_R(z_I) + B \, ,
\label{eq:shiftedAmplitude}
\ee
where $\hat{\mathcal{A}}_{L,R}(z_I)$ are the two shifted lower point amplitudes that appear on either side of the factorization on the pole $z_I$, and $B$ is a boundary term coming from the $z\to\infty$ region. 
Therefore, under a shift where $B=0$, the calculation of tree-level on-shell amplitudes is reduced to finding solutions to the linear equations $\hat{P}^2_i=0$, which accounts for all possible factorization channels, where the residues of the shifted amplitude are lower point on-shell amplitudes.  The shifted amplitude is therefore determined by combining lower point on-shell amplitudes connected by scalar propagators.

Note that condition (3) forces that there must be at least two shifted momenta. This minimal option leads to the BCFW shift~\cite{Britto:2005fq}.  Nonminimal options include the shift of all the external lines, as explored in~\cite{Elvang:2008vz,Cohen:2010mi}, which is also known as the CSW shift~\cite{Cachazo:2004kj}. 

\subsection{Recursion Relations for Form Factors}
\label{sec:recrelationFFs}

We now apply the same ideas to the case of form factors.
In a scattering amplitude with all momenta outgoing, the sum of their momenta vanishes by momentum conservation. However, in a form factor $\mathcal{F}^\mathcal{O}_\eta(\alpha)$ the operator $\mathcal{O}$ injects an arbitrary momentum, so the sum of momenta is nonvanishing.  Therefore, when building a recursion relation for form factors, condition (3) in \cref{eq:amponshellconditions} should be dropped.  This implies that it is possible to analytically continue the form factor by considering a single line shift.
Moreover, since the tree-level form factors include Wilson lines, they contain ``eikonal poles'' when gluon momenta are collinear to the Wilson line $\sim 1/\eta\cdot q$, in addition to the normal poles from on-shell propagators. 

When defining a shifted form factor, we can eliminate the would be additional $z$ dependence from the eikonal poles by imposing the condition $\eta \cdot r_i=0$ for each $i$. 
So we define a shift in the external momenta $p^\mu_i$ of a form factor with a Wilson line in the direction $\eta^\mu$ such that $\hat{p}^\mu_i= p^\mu_i + z\s r_i^\mu$ with the conditions:

\clearpage

\begin{center}
Momentum shift for form factors
\end{center}
\vspace{-20pt}
\begin{align}
(1) \quad & \quad r_i\cdot r_j = 0 \qquad\text{for any pair, including } i = j\,;\nn \\[3pt]
(2) \quad & \quad p_i\cdot r_i = 0\qquad \text{for all } i\,; \label{eq:ffrrconditions} \\[3pt]
(3) \quad & \quad \eta\cdot r_i = 0\qquad \text{ for all } i\,.\nn
\end{align}
Conditions (1) and (2) serve the same purpose as in the case of amplitudes.  Enforcing condition (3) avoids having $z$-dependent eikonal poles; the only poles in form factors are the ones coming from intermediate particles going on-shell.  We can again use the Cauchy theorem to write 
\be
\mathcal{F}^{\mathcal{O}}_{\eta} = \sum_{z_I} \hat{\mathcal{A}}(z_I)\frac{1}{P^2_I}\hat{\mathcal{F}}^{\mathcal{O}}_{\eta}(z_I) + B_{\eta}\,,
\ee
where the $B_{\eta}$ term corresponds to the contribution from $z\to\infty$.  If $B_{\eta}$ vanishes, the form factor $\mathcal{F}^{\mathcal{O}}_{\eta} \equiv \hat{\mathcal{F}}^{\mathcal{O}}_{\eta}(z=0)$ can be computed using the shifted lower point form factor $\hat{\mathcal{F}}^{\mathcal{O}}_{\eta}$ and a shifted on-shell amplitude $\hat{\mathcal{A}}(z_I)$ connected by a scalar propagator.  Note that the shifted momenta must always be connected to the scattering amplitude side, due to momentum conservation. Another way to see this is that a shift of a momentum connected to the form factor would never lead to a cut propagator with non-trivial $z$-dependence.  

We will show that $B_{\eta} = 0$ for gauge theories with scalar and/or fermionic matter for the appropriate choice of momentum shift.  Then we only have the factorizable contributions, and the recursion relation can be represented diagrammatically:

\begin{center}
	\includegraphics[width=0.3\linewidth]{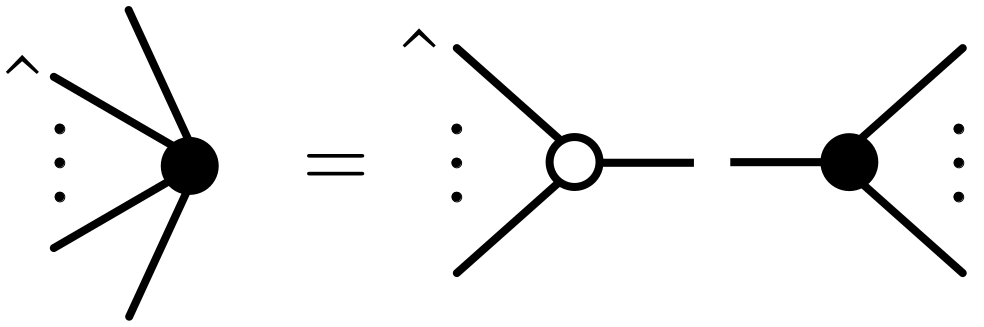}
\end{center}
where the black vertices denote Wilson line form factors and the white vertices denote on-shell amplitudes.
The minimal choice that satisfies the conditions in \cref{eq:ffrrconditions} is given by a single line shift, where only one of the $r^\mu_i$ is non-vanishing. 
The hat $(\,\hat{}\,)\,$ in the diagram denotes the shifted momenta, which necessarily belongs to the on-shell amplitude in the right hand side.
Using spinor helicity variables, this is either an 
$|\hat{i}\rangle$-shift 
\begin{subequations}
\be
|\hat{i}\rangle = |i\rangle - z |\eta\rangle\,,
\ee
or an $|\hat{i}]$-shift
\be
|\hat{i}] = |i] - z |\eta]\,,
\ee
\end{subequations}
where $|\eta\rangle$ or $|\eta]$ are the spinors in \cref{eq:etaSpinors} associated with the null direction of the Wilson line.  The choice of shift depends on the behavior at infinity of the particular form factor of interest, as we explore in the next section.

\subsection{Large-$z$ Behavior and Constructibility}

\begin{figure}[h!]
	\centering
	\vspace{10pt}
	\includegraphics[width=0.4\linewidth]{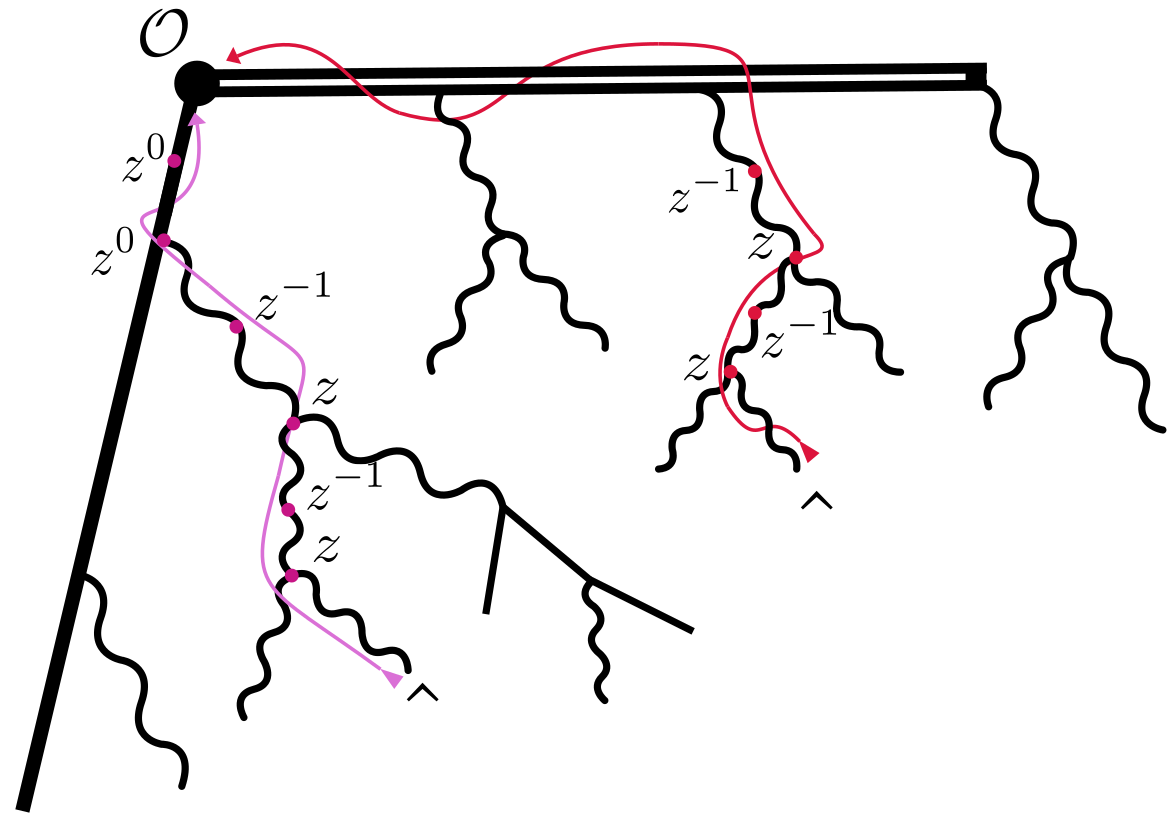}
	\caption{Here we show a characteristic example of a Feynman diagram that appears in the form factors.  There are used as inputs for the characterization of the large-$z$ behavior of the form factors. The dot denotes the operator insertion, and the double line denotes the eikonal factors coming from the perturbative expansion of the Wilson line. Shifting an external gluon injects a $z$-dependence into the momentum that flows from the gluon to the operator insertion, necessarily going through either the charged matter line (single-line shift corresponding to the pink flow) or the Wilson line (single-line shift corresponding to the red flow).}
	\label{fig:shiftfeynmandiagram}
\end{figure}

Now we discuss the $z\to \infty$ limit of the helicity form factors.  Our goal is to derive conditions for when the boundary term $B_\eta$ in \cref{eq:shiftedAmplitude} vanishes so that the form factors are constructible in terms of lower point inputs. This argument is based on considering the scaling behavior of all possible Feynman diagrams that can appear, following the approach from \cite{Britto:2005fq}.  An example diagram is drawn in \cref{fig:shiftfeynmandiagram}.

A generic Feynman diagram for a form factor is a combination of propagators, vertices, eikonal propagators, eikonal vertices, external polarizations, and momenta insertions.
When shifting the momenta of an external leg, the $z$-dependence flows from this leg to the operator insertion going through $n$ vertices and $n$ propagators.
Since the shift is proportional to $\eta$, eikonal propagators and eikonal vertices do not depend on $z$.
Gauge boson and scalar propagators scale as $1/z$, while fermion propagators scale at worst as $z^0$. The fermion-gauge boson vertex scales as $z^0$, the gauge boson 4-point self-interaction scales as $z^0$, and the gauge boson 3-point self-interaction and the scalar-gauge boson vertex scale at worst as $z$.
Therefore, the worst large-$z$ scaling of a diagram due to the presence of vertices and propagators is $z^0$.

The only wavefunction that can scale with $z$ is the one corresponding to the particle whose momentum is being shifted.
If the particle is a fermion, the best possible scaling of the associated wavefunction is $z^0$, when one chooses to shift the spinor that has the opposite helicity of the fermion wavefunction.
If the particle is a gauge boson, it is possible to always choose a shift such that the polarization vector scales as $z^{-1}$. For example, the wavefunction for a negative-helicity vector with momentum $p^\mu$ is proportional to $ \frac{\ra{p}\ls{\xi}}{\sq{\xi p}}$, so that a $\rs{\hat{p}}$-shift ensures a $z^{-1}$ scaling.

Next, we discuss the $z$-dependence associated with the different operators.
As emphasized above, we only consider the operators listed in \cref{eq:treeformfactors}, which are associated with the leading behavior in the collinear limit. 
We leave the study of the convergence of form factors associated with subleading corrections~\cite{Beneke:2002ni,Freedman:2011kj,Bhattacharya:2018vph} for future work.

\subsubsection{Scalar and Fermionic Operators} 
Consider the case where the operator $\mathcal{O}$ corresponds to a scalar or a fermion. The $B_\eta$ factors vanish for both scalar and fermionic form factors if one shifts an external gauge boson.  In the case where the shifted particle is a gauge boson of a fixed polarization, one is free to choose either an $|\hat{i}\rangle$-shift or a $|\hat{i}]$-shift since the full diagram scales as $1/z$.

Next, we consider the case where the external state does not contain gauge bosons and only fermions are present.
If the shifted fermion is connected to an internal gluon, this leads to diagrams that can scale at most as $1/z$. 
The only potentially problematic $\sim z^0$ contribution comes from the case where the shifted fermion connects directly with the operator insertion. However, one can always avoid such a contribution by shifting the fermion with a different chirality than that of the operator.  We conclude that for the case where there is no external gauge boson, then one can shift an external fermion with the opposite helicity as compared to the operator insertion.  Finally, we note that the scalar case with no external gauge bosons is not constructible using a 1-line shift.

\subsubsection{Gauge Boson Operators}
We now consider the case where the operator insertion corresponds to the one interpolating a non-Abelian gauge boson, see Eqs.~(\ref{eq:gluonFFdef}). We argue that the $B_\eta$ terms vanish if one shifts an external gauge boson line with the opposite helicity as compared to that of the operator insertion.

Consider the case where one shifts a gauge boson line, such that corresponding shifted wavefunction scales as $1/z$. Then the combination of vertices, propagators and wavefunction scales at worst as $1/z$. However, contrary to the fermion case, now the operator insertion contains a momentum and potentially scales as $z$. Under a $|\hat{i}\rangle$-shift, the left-chiral part of the momentum proportional to $z$ that runs through the operator is proportional to $\la{\eta}$. Therefore, the projection along $\eta_\perp$ kills this contribution.  Note that if one projected along the $\eta_\perp^*$ direction, this form factor would grow with $z$, see Eqs.~(\ref{eq:etaslashed}) and (\ref{eq:gluonFFdef}). 
We therefore conclude that the $B_\eta$ terms only vanish if one shifts a gauge boson with the opposite helicity from the gauge boson interpolated by the operator.

This argument implies that we can not discard the possibility of having a $z^0$ contribution in the case where the form factor has an external state where all of the gauge bosons have the same helicity as the one coming from the operator. For such cases, a two-line shift is required. However, as it will be discussed in \cref{sec:GBFormFactorsQCD} below, this case is also constructible from single-line shifts. The physical reason is that the minimal form factor with a single gluon is chiral, in the sense that only its left or right chiral part is dynamical, and the rest is fixed by the Wilson line.

If there are only fermions in the external state, shifting the chirality of an external momentum corresponding to the same chirality of the operator ensures that the diagram scales as $1/z$. Since there is always a fermion whose wavefunction will not scale under the shift, such form factors are always constructible.

\subsection{Recursion for Massive Wilson lines}
\label{sec:incmasses}

So far we have assumed that the quarks are massless and the Wilson lines correspond to null directions. It is straightforward to extend the formalism to the case where the Wilson line corresponds to a massive trajectory, so $\eta^2\neq 0$. Using the massive spinor formalism \cite{Arkani-Hamed:2017jhn}, one can write $\slashed \eta = \ra{\eta^I}\ls{\eta_I}$, with ${}^I$ being a little group index. Then, the shift
\be
\ra{\hat{p}} \,=\, \ra{p} \,+\, z\, \slashed \eta \rs{p}
\label{eq:massiveshift}
\ee
leaves the massless momenta $\hat{p}^\mu$ on-shell, $\hat{p}^2=0$, and yields no $z$-dependence from eikonal propagators, $\eta\cdot\hat{p}=\eta\cdot p$ since $\ls{p}\slashed\eta\slashed\eta\rs{p}=\eta^2\sq{pp}=0$.
All the conclusions from above then follow.

In the case of QED and QCD with massive matter, shifting an external photon or gluon leads to a convergent recursion relation, the reason being that the $z$ scaling in the $z\to\infty$ limit is not changed by the presence of masses.
However, shifting massive fermions becomes nontrivial. Development of these techniques for this case would allow to cross check and push the state-of-the art calculation of splitting functions with massive partons \cite{Dhani:2023uxu,Craft:2023aew}. 
In the case of a spontaneously broken gauge theory, it would be interesting to explore the connections with electroweak radiation 
\cite{Garosi:2023bvq,Nardi:2024tce}.

\clearpage

\section{QED}
\label{sec:QED}
In this section, we apply the recursion relations above to derive tree-level helicity form factors in QED, a theory with scalar or fermionic matter coupled to an Abelian gauge boson (a photon).  We will emphasize some universal features that emerge and will present an organization in terms of ``minimal helicity violating'' form factors, in close analogy with the well known properties of helicity amplitudes.

\subsection{MHV Form Factors in Scalar QED}
In the case of a scalar, the free theory wavefunction is trivial, so the form factor is $\mathcal{F}^\phi_\eta(\phi_1)=1$, see \cref{eq:scalarWF}.  At first order in the gauge coupling the operator overlaps with a state containing a scalar and a photon.  If this photon has positive helicity, the form factor $\mathcal{F}^\phi_\eta(\phi_1 \gamma_2^+)$ can be computed from the recursion relation
\be
\includegraphics[width=0.3\linewidth, valign=c]{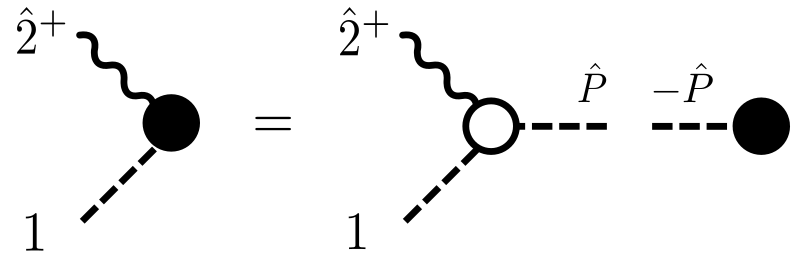}
\,\quad
=\,\quad\frac{1}{\ab{12}}\frac{\ab{1\eta}}{\ab{2\eta}}\,.
\label{eq:FFscalar12}
\ee
It is interesting that this result has an explicit dependence on the $\eta^\mu$ direction, even though the building blocks entering the recursive calculation did not.  In other words, the presence of the Wilson line is \emph{emergent} from the recursive point of view.

The generic form factor with an arbitrary number of same-helicity photons in the final state can be computed recursively as follows. Using the fact that the on-shell amplitude with two scalars and $n$ same-helicity photons vanishes for $n\geq 2$, only the three point amplitude appears in the recursive calculation.  If we shift $\ra{\hat{2}}$ in $\mathcal{F}^\phi_\eta(\phi_1 \gamma_2^+\dots \gamma_n^+)$, the only contribution comes from the three point amplitude and the form factor with $n-1$ same-helicity photons.  Using induction, we find that the scalar QED tree-level form factor $\mathcal{F}^\phi_\eta(\phi_1 \gamma_2^+\dots \gamma_n^+)$ is given by 
\begin{align}
\includegraphics[width=0.35\linewidth, valign=c]{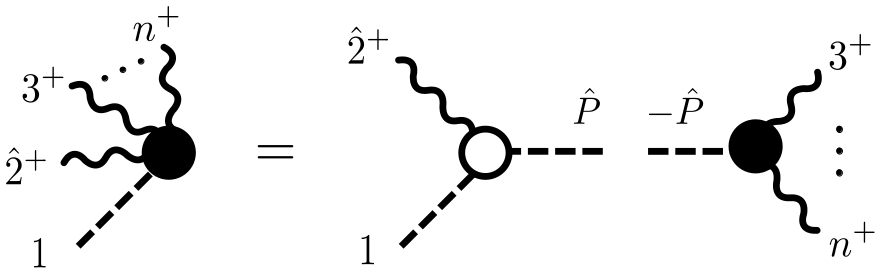}
\quad &= \quad
\frac{\sq{12}\sq{2\hat{P}}}{\sq{1\hat{P}}}\frac{1}{\ab{12}\sq{21}}\frac{\ab{\hat{P} \eta }^{n-2}}{\ab{\hat{P} 3}\ab{3\eta}\dots \ab{\hat{P} n}\ab{n\eta}}\notag\\
&=\quad 
\frac{\ab{\eta 1}^{n-1}}{\ab{1 2}\ab{2\eta}\dots \ab{1 n}\ab{n\eta}}\,.
\label{eq:QEDscalarMHV}
\end{align}
The second line is obtained by performing the usual manipulations for these kinds of calculations. The term
$\sq{2\hat{P}}/\sq{1\hat{P}}$ can be multiplied by $\ab{\hat{P}\eta}/\ab{\hat{P}\eta}$, and using momentum conservation and the fact that $\ra{\eta}$ annihilates the $z$-dependent part of the momentum, one obtains $\sq{2\hat{P}}/\sq{1\hat{P}}=-\ab{1\eta}/\ab{2\eta}$. For the terms of the type $\ab{\hat{P}\eta}/\ab{\hat{P}k}$, one can multiply and divide by $\sq{\hat{P}1}$ to get $\ab{\hat{P}\eta}/\ab{\hat{P}k}=\ab{\eta 2}/\ab{k \hat{2}}$. This last term depends on $z$, which is however fixed by the on-shell condition $\hat{P}^2=\ab{1\hat{2}}\sq{21}=0$, which implies $0=\ab{1\hat{2}}=\ab{12}-z\ab{1\eta}$, and therefore $z=\ab{12}/\ab{1\eta}$. Using this, one gets the last line of \cref{eq:QEDscalarMHV}.

We will refer to this form factor as the ``Maximally Helicity Violating'' (MHV) form factor, in analogy with the common terminology in the scattering amplitudes literature.  As in the case of scattering amplitudes, it is expressed as a holomorphic function that only depends on the angle spinors.  A similar calculation yields the $\overline{\text{MHV}}$ form factor
\begin{align}
\mathcal{F}^\phi_\eta(\phi_1 \gamma_2^-\dots \gamma_n^-) = \frac{\sq{\eta 1}^{n-1}}{\sq{1 2}\sq{2\eta}\dots \sq{1 n}\sq{n\eta}}\,,
\label{eq:QEDscalarMHVBar}
\end{align}
which is an anti-holomorphic function that only depends on the square spinors.  Note that contrary to the case of scattering amplitudes, a form factor where all vectors have the same helicity is non-vanishing, so the MHV classification starts with this configuration.  

\subsection{MHV Form Factors and Soft Radiation}
\label{sec:SoftExchanges}
We will now explain that the MHV form factors are entirely determined by soft photon
exchanges.
To show this, we begin by identifying a connection between the MHV form factor and the form factor with two Wilson lines.  Taking the limit of infinite energy for the scalar, $\lim_{\omega\to\infty} p_1^\mu$ with $p_1^\mu=\omega\s\rho^\mu$, corresponds to the eikonal limit for the scalar such that it becomes a recoilless probe.  In this limit, the scalar becomes a Wilson line in the direction $\rho^\mu$, and the form factor computed above should reproduce a form factor for two Wilson lines, which we will denote $\mathcal{F}_{\eta\rho}$. Indeed, the latter is given by 
\be 
\mathcal{F}_{\eta\rho}(\gamma_2^+\dots \gamma_n^+)
=
\lim_{\omega\to\infty} \mathcal{F}_{\eta}^\phi(\phi_1\,\gamma_2^+\dots \gamma_n^+)\vert_{p_1^\mu=\omega\s \rho^\mu}
=
\frac{\ab{\eta\rho}^{n-1}}{\ab{\rho 2}\ab{2\eta}\dots \ab{\rho n}\ab{n \eta}} \, ,
\label{eq:FFeikonalMHV}
\ee
where $\mathcal{F}_{\eta\rho}(\gamma_2^+\dots \gamma_n^+) = \la{\gamma_2^+ \dots \gamma_n^+}W_\eta^\dag(0) W_\rho(0) \ra{0}$.
The fact that this result takes the same form as \cref{eq:QEDscalarMHV} demonstrates 
that the scalar in the MHV form factor acts as a Wilson line.  Since the scalar has infinite momentum, any radiated photons are soft since they have finite momentum.  We can check this explicitly as follows.

There is a suggestive way to write the result in \cref{eq:QEDscalarMHV} that makes clear that the MHV form factor contains purely soft interactions. We can identify \cref{eq:FFscalar12} with the eikonal factor~\cite{Dixon:2013uaa} 
\be
\mathcal{S}^{ab}_k \equiv \frac{\ab{ab}}{\ab{ak}\ab{kb}}\,,
\ee
which captures the soft limit of a photon being emitted by a pair of eikonal lines \cite{PhysRev.140.B516}.
Then we can rewrite the MHV form factor to show that it is given by the product of soft factors:
\be
\includegraphics[width=0.2\linewidth, valign=c]{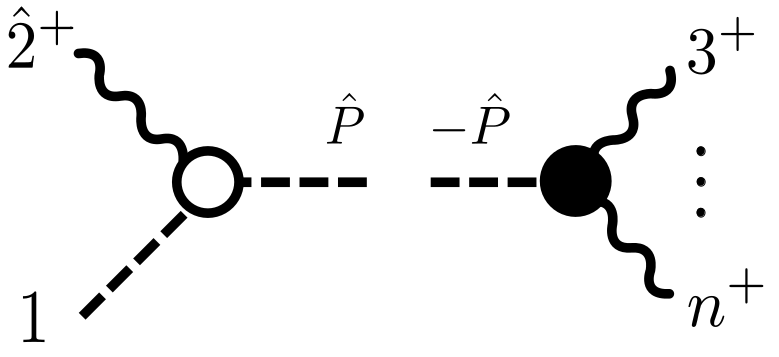}
\,\,\,=\,\,\,
\frac{\sq{12}\sq{2\hat{P}}}{\sq{1\hat{P}}}\frac{1}{\ab{12}\sq{21}}\prod_{i=2}^n \mathcal{S}^{\hat{P}\eta}_k\,.
\ee
The explicit spinor factors in this expression are equivalent to $\mathcal{S}^{1\eta}_2$, while the rest of the soft factors reduce to $\mathcal{S}^{\hat{P}\eta}_k=\mathcal{S}^{1\eta}_k$, which leads to the MHV result in \cref{eq:QEDscalarMHV} above. This shows that the soft structure is correctly reproduced by the recursion relation, and indeed one obtains the eikonal form in \cref{eq:FFeikonalMHV}.  This is in contrast with the non-MHV form factors, which cannot be expressed only in terms of soft factors, as we will see below.

The relevance of this observation is that jet functions are defined as a ratio between the Wilson line form factors of the type in \cref{eq:FormFactorDef} and the form factors where the operator is replaced by a Wilson line in the direction of motion of the hard particle in the jet \cite{Agarwal:2021ais}. Therefore, the jet functions are trivial in the MHV sector.  To our knowledge this is a novel observation, whose consequences we will explore in future work.

\subsection{MHV Form Factors in Fermion QED}

Now we turn to the case where the charged matter is composed of fermions.  The logic follows in close analogy with the scalar QED case.  The form factors that describe the leading effects of collinear and soft emission of external particles emerge from the free wavefunction, which in the fermion case is either $\ra{p}$ or $\rs{p}$, depending on the fermion helicity, see \cref{eq:fermionWF}.

The main difference with respect to the scalar case is that now there are two distinct form factors, one for each helicity of the fermion. Both can be computed using the free fermion wavefunction and the three point amplitude.  This calculation yields the following form for the MHV form factor,
\be
\mathcal{F}^{f^-}_{\eta} (f_1^- \gamma_2^+)= \frac{\ab{\eta 1}\la{1}}{\ab{12}\ab{2\eta}}\ra{\eta}\,,
\label{eq:2pointfermionFFmp}
\ee
and the following for the next-to-MHV or NMHV form factor,
\be
\mathcal{F}^{f^-}_{\eta} (f_1^- \gamma_2^-)= \frac{\ls{\eta}\slashed q}{\sq{12}\sq{2 \eta}}\ra{\eta} \,,\label{eq:2pointfermionFFmm}
\ee
where we have defined the momentum $q^\mu$ as the sum of all the momenta in the form factor, in this case $q^\mu = p_1^\mu+p_2^\mu$.
In writing these formulas, we have made the contraction with the projection $\ra{\eta}$ explicit.  This emphasizes that the first form factor is proportional to the fermion wavefunction and therefore it can be thought of as modifying the scalar result above.
The nomenclature chosen indicates that the MHV form factor is a holomorphic function of the angle spinors. The NMHV form factor is not holomorphic due to the presence of $\ls{\eta}\slashed q$.

The fermion form factor with any number of photon legs with the same helicity can be computed following the same approach as the scalar case presented above.
The MHV form factor is
\be
\mathcal{F}^{f^-}_{\eta} (f_1^- \gamma_2^+\dots \gamma_n^+)
=
\frac{\ab{1\eta}^{n}}{\ab{12}\ab{2\eta}\cdots \ab{1n}\ab{n\eta}}\,.
\ee
The N$^{n-1}$MHV form factor for the same fermion helicity and with all photon helicities flipped can also be computed for any number of photons. We find
\be
\mathcal{F}^{f^-}_{\eta} (f_1^- \gamma_2^-\dots \gamma_n^-)
=
\frac{\sq{1\eta}^{n-2}\ls{\eta}q\ra{\eta}}{\sq{12}\sq{2n}\cdots \sq{1n}\sq{n\eta}} \,.
\ee
Again, notice that while the MHV form factor is holomorphic, the N$^{n-1}$MHV form factor is not. 
However, the N$^{n-1}$MHV form factor does have a simple form because the $\overline{\text{MHV}}$ amplitude appears as a building block in its derivation.

Both fermionic MHV and N$^{n-1}$MHV form factors are related to the scalar MHV form factor by a simple kinematic prefactor
\begin{subequations}
\bea
\mathcal{F}^{f^-}_{\eta} (f_1^- \gamma_2^+\dots \gamma_n^+)
&=&
\mathcal{F}^{\phi}_{\eta} (\phi_1 \gamma_2^+\dots \gamma_n^+)\, \ab{1\eta} \, ,\\[4pt]
\mathcal{F}^{f^-}_{\eta} (f_1^- \gamma_2^-\dots \gamma_n^-)
&=&\mathcal{F}^{\phi}_{\eta} (\phi_1 \gamma_2^-\dots \gamma_n^-)\, \frac{\ls{\eta}q\ra{\eta}}{\sq{1\eta}} \,,
\label{eq:fermiontoscalarFFQED}
\eea
\end{subequations}
where $\mathcal{F}^{\phi}_{\eta} (\phi_1 \gamma_2^+\dots \gamma_n^+)$ and $\mathcal{F}^{\phi}_{\eta} (\phi_1 \gamma_2^-\dots \gamma_n^-)$ are given by \cref{eq:QEDscalarMHV,eq:QEDscalarMHVBar} respectively.
In fact, notice that the fermionic MHV form factor is the scalar one, times the projected fermion wavefunction $\ab{1\eta}$. This makes the connection between the MHV form factor and the soft form factor explicit, following the discussion in \cref{sec:SoftExchanges}.
Indeed, the MHV form factor is entirely determined by eikonal gauge bosons interactions, as the only difference between the fermion and scalar form factors is the extra factor of the fermion's projected wavefunction.

\bigskip

\subsection{NMHV Form Factors}

We now turn to the calculation of the NMHV form factors.  We begin with $\mathcal{F}^\phi_\eta(\phi_1 \gamma_2^- \gamma_3^+)$ in scalar QED.
We can compute the form factor recursively using either a $\rs{\hat{2}}$-line shift or a $\ra{\hat{3}}$-line shift.
Shifting $\rs{\hat{2}}$, we get two contributions:
\hspace{.3cm}
\begin{equation}
	\includegraphics[width=0.33\linewidth, valign=c]{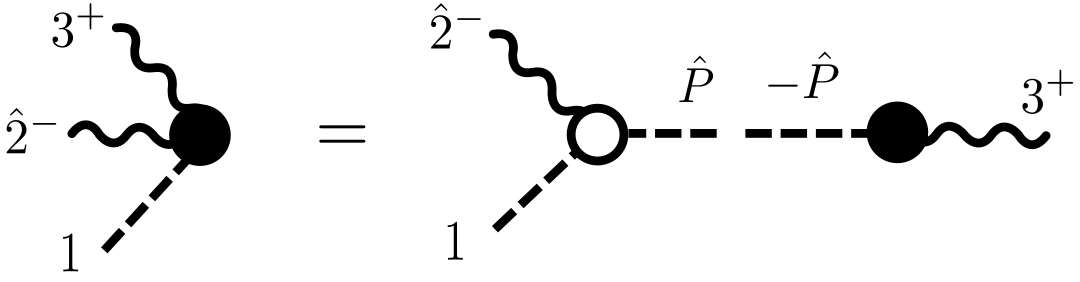}
\,\,+\,\,
	\includegraphics[width=0.16\linewidth, valign=c]{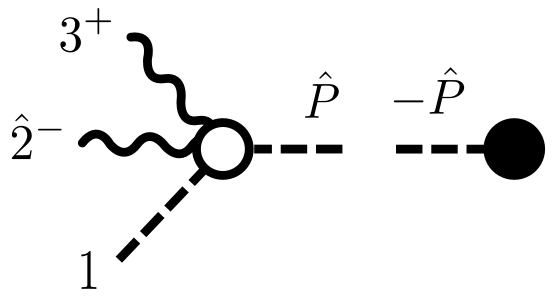}
	\label{eq:NHMVdiagrams}
\end{equation}
\hspace{.3cm}

The sum of both terms gives
\be
\mathcal{F}^\phi_\eta(\phi_1 \gamma_2^- \gamma_3^+)
=
\frac{\ab{12}}{\la{3}q\rs{\eta}}\bigg(\frac{\sq{1\eta}\la{\eta}1+2\rs{\eta}}{\ab{\eta 3}\sq{2\eta}s_{12}}
+
\frac{\la{2}q\rs{\eta}}{\ab{31} s_{123}}\bigg)\,,
\label{eq:FFscalarNMHV}
\ee
where the first term comes from the first diagram that involves the 3-point amplitude, and the second term with the pole at $s_{123}=(p_1+p_2+p_3)^2= q^2$ comes from the second diagram which involves the MHV amplitude and the scalar wavefunction.
Note that if we had performed a shift on $\ra{\hat{3}}$, we would obtain a very different (but equivalent) expression. This, together with the appearance of the nonlocal poles of the type $\la{3}q\rs{\eta}$, is reminiscent of the features of NMHV amplitudes \cite{Hodges:2009hk}. It would be therefore interesting to explore a formulation of the Wilson line form factors akin to the Grassmannian perspective of scattering amplitudes, see \emph{e.g.}~\cite{Arkani-Hamed:2009ljj}.

Now consider the case of fermion QED. There is a single NMHV form factor:
\be
\mathcal{F}^{f^-}_\eta(f^-_1 \gamma_2^- \gamma_3^+)
=
\frac{\ab{12}}{\la{3}q\rs{\eta}}\left(\frac{\la{\eta}1+2\rs{\eta}}{\sq{1\eta}}\frac{\sq{1\eta}\la{\eta}1+2\rs{\eta}}{\ab{\eta 3}\sq{2\eta}s_{12}}
+
\frac{\ab{12}\la{\eta}q\rs{\eta}}{\la{2}q\rs{\eta}}\frac{\la{2}q\rs{\eta}}{\ab{31} s_{123}}\right)\,.
\ee
This form is expected due to the relations between MHV scalar and fermion form factors, and MHV amplitudes between fermions and scalars. The relations between MHV form factors are given above, while the relations among amplitudes are given by 
\begin{align}
\mathcal{A}(f^-\bar{f}^+\gamma^+\gamma^-)=\frac{\ab{14}}{\ab{24}}\mathcal{A}(\phi\phi\gamma^+\gamma^-) \,,
\end{align}
etc.  Applying these relations recursively gives
\begin{subequations}
\begin{align}
\hspace{-8pt}\mathcal{F}^{f^-}_\eta(f_1^-\gamma_2^-\gamma_3^+)_\text{diag 1}
&=
\frac{\ab{12}}{\ab{\hat{P}2}}\ab{\hat{P}\eta}\mathcal{F}^\phi_\eta(\phi_1 \gamma_2^- \gamma_3^+)_\text{diag 1}
=
\frac{\la{\eta}1+2\rs{\eta}}{\sq{1\eta}}\mathcal{F}^\phi_\eta(\phi_1 \gamma_2^- \gamma_3^+)_\text{diag 1}\,,\\[3pt]
\hspace{-8pt}\mathcal{F}^{f^-}_\eta(f_1^-\gamma_2^-\gamma_3^+)_\text{diag 2}
&=
\frac{\ab{12}}{\ab{\hat{P}2}}\ab{\hat{P}\eta}\mathcal{F}^\phi_\eta(\phi_1 \gamma_2^- \gamma_3^+)_\text{diag 2} 
=
\frac{\ab{12}\la{\eta}q\rs{\eta}}{\la{2}q\rs{\eta}}\mathcal{F}^\phi_\eta(\phi_1 \gamma_2^- \gamma_3^+)_\text{diag 2}\,,
\end{align}
\end{subequations}
where the ``diag'' subscript refers to the two recursive diagrams that appear on the right hand side of \cref{eq:NHMVdiagrams}. We see that the MHV relations between scalar and fermion form factors and amplitudes holds individually for the diagrams that appear in the recursion relations. However, the diagrams are evaluated on different kinematic poles, which leads to a non-homogeneous relation between scalar and fermion form factor at the NMHV level.

We can also study the eikonal limit of the NMHV form factor.  We can explore this by taking the limit as in \cref{eq:FFeikonalMHV} above $\lim_{\omega\to\infty} p_1^\mu$ for the scalar momenta $p_1^\mu=\omega\s\rho^\mu$
Applying this limit to \cref{eq:FFscalarNMHV}, we find
\be
\mathcal{F}_{\eta\rho}(\gamma_2^- \gamma_3^+)
=
\lim_{\omega\to\infty} \mathcal{F}_{\eta}^\phi(\phi_1\,\gamma_2^- \gamma_3^+)\vert_{p_1^\mu=\omega \rho^\mu}
=
\frac{\sq{\rho\eta}\ab{\eta\rho}}{\sq{\eta 2}\sq{2 \rho}\ab{\eta 3}\ab{3\rho}}\,.
\label{eq:FFeikonalNMHV}
\ee
Returning to the diagrams in \cref{eq:NHMVdiagrams}, one can see that the second diagram goes to zero in this limit, so that the only contribution comes from the first diagram that involves the 3-point amplitude. 
In fact, this pattern continues. 
In this limit there is always a contribution to the form factor that scales as $\omega^0$. The on-shell propagator scales as $1/\omega$, so the only diagram that can have a finite contribution is one that involves an amplitude that scales as $\omega$. Because of helicity scaling of the external wavefunctions, all amplitudes scale as $\omega^0$, except for the three point amplitude. 
Indeed, the three point amplitude $\mathcal{A}(\phi_1 \phi^*_2 \gamma^-_3)=\ab{13}\ab{32}/\ab{12}$ scales as $\omega^0$ (only scaling the energy dependence of the spinors), but in the limit of interest the two scalars become parallel and therefore $\ab{12}\to 0$ as $\omega \to \infty$.\footnote{The special kinematics of the three point amplitude allows us to set all square spinors proportional to each other, in particular $\ls{2}= r \ls{1}$ for some constant $r$. 
In the limit of interest, the two scalars become collinear and therefore $p_\phi+p_{\phi^*}+p_\gamma\to p_\phi+p_{\phi^*} = (\ra{1}+r \ra{2})\ls{1}=0$, which implies $\ab{12}\to 0$.} This divergence compensates the $1/\omega$ scaling of the propagator, resulting in a finite contribution to the form factor.

This observation allows us to find a general expression for the tree level radiation
\be
\mathcal{F}_{\eta\rho}(\gamma_1^-\cdots \gamma_i^- \gamma_{(i+1)}^+\cdots \gamma_n^+) 
\,=\,\prod_{j=1}^i \frac{\sq{\rho\eta}}{\sq{j\eta}\sq{\rho j}}\prod_{k=i+1}^n \frac{\ab{\rho\eta}}{\ab{k\eta}\ab{\rho k}}\,.
\ee
Interestingly, it is the special properties of complexified three particle kinematics that allows us to find a general expression for this form factor in this limit.  This is a very different approach from the canonical argument that relies on the universal eikonal factor and a sum over permutations.

\subsection{Form Factors with Massive Wilson Lines}
As discussed in \cref{sec:incmasses}, it is straightforward to consider the case where the Wilson lines are tilted inside the lightcone and represent massive trajectories. 
The calculation of the form factor $\mathcal{F}^\phi_\eta(\phi \gamma^-)$ is very similar to the one in \cref{eq:FFscalar12}, but with the difference that now the shift is of the type in \cref{eq:massiveshift}, $\rs{\hat{2}}=\rs{2}+z\slashed \eta\ra{2}$. Therefore, the on-shell condition implies $z=-\sq{12}/\la{2}\eta\rs{1}$. In this case the form factor is given by
\be
\mathcal{F}^\phi_\eta(\phi_1 \gamma_2^-)
\,=\,
\frac{1}{\sq{12}}\frac{\la{2}\eta\rs{1}}{\la{2}\eta\rs{2}} \,.
\ee
By sending $\eta^\mu$ to the lightcone, so $\slashed \eta = \ra{\eta}\ls{\eta}$, this result reproduces the expression in \cref{eq:FFscalar12} as it must.

\section{QCD}
\label{sec:QCD}
In this section, we explore the application of the above ideas to QCD, non-Abelian gauge theories with scalar or fermion matter in the fundamental representation.  The case of QCD is very similar to QED, except now we must keep track of color. In the following, the operator $\mathcal{O}$ that appears in the form factor is either a scalar or fermion in the fundamental representation or a vector operator transforming in the adjoint. At each order in perturbation theory, the form factor will depend explicitly on the color index associated with the operator. The physical picture is that the form factor arises as part of a factorized amplitude, where the hard interaction produces a particle with a given spin, momentum, and $SU(N)$ quantum number.

\subsection{Color Ordering}
As in the case of the amplitudes, it is very convenient to organize the calculation in terms of ``color ordered'' form factors. 
In the case of the scalar form factor, the color decomposition is given by
\be
\mathcal{F}^{\phi_j}_\eta\big(\phi_i g^{h_2}_{A_2} g^{h_3}_{A_3} \dots g^{h_n}_{A_n}\big)
\,=\,
\sum_{\sigma\in S_{n-1}} \big(T^{A_{\sigma_2}}T^{A_{\sigma_3}}\dots T^{A_{\sigma_n}}\big)_{ij}\,\mathcal{F}^{\phi}_\eta\big[\phi g^{h_{\sigma_2}}_{\sigma_2} g^{h_{\sigma_3}}_{\sigma_3} \dots g^{h_{\sigma_n}}_{\sigma_n}\big]\,,
\ee
where $i$ is the color index of the scalar in the state, $j$ is the color index of the scalar field in the operator, $T^{A_i}$ is the generator associated with the $i^\text{th}$ gluon, and $S_n$ is the permutation group of $n$ elements.
On the right hand side, the square brackets denote the color-ordered form factor, which carries only kinematic information.
Similarly, the gluon form factors are given by
\be
\mathcal{F}^{g^{h}_{A}}_\eta\big(g^{h_1}_{A_1} g^{h_2}_{A_2} \dots g^{h_n}_{A_n}\big)
\,=\,
\sum_{\sigma\in S_n} \text{tr}\big(T^{A_{\sigma_1}}T^{A_{\sigma_2}}\dots T^{A_{\sigma_n}} T^A\big) \mathcal{F}^{g^{h}}_\eta\big[ g^{h_{\sigma_1}}_{{\sigma_1}} g^{h_{\sigma_2}}_{{\sigma_2}} \dots g^{h_{\sigma_n}}_{{\sigma_n}}\big]\,,
\ee
where we used the cyclic invariance of the trace to fix the last entry, which corresponds to the generator associated with the operator $\mathcal{O}$.

This basis of color-ordered form factors is redundant. As in the case of scattering amplitudes, color ordered form factors obey a set of identities known as $U(1)$ decoupling identities \cite{Berends:1987cv,Mangano:1987xk,Berends:1987me}, which can be obtained by imposing that the amplitude for a photon and $n-1$ gluons vanishes.  This must be the case since the photon generator commutes with all other generators, which implies that the coupling (which is proportional to the associated structure constant) vanishes. By writing $T^A=\mathbb{1}$, all factors with identical traces should vanish independently:
\be
0\,=\,
\mathcal{F}^{g^{h}}_\eta\big[ g^{h_{1}}_{1} g^{h_{2}}_{2} \dots g^{h_{n}}_{n}\big]
+
\mathcal{F}^{g^{h}}_\eta\big[ g^{h_{2}}_{2} \dots g^{h_{n}}_{n} g^{h_{1}}_{1}\big]
+
\dots
+
\mathcal{F}^{g^{h}}_\eta\big[  g^{h_{n}}_{n} \dots  g^{h_{1}}_{1}g^{h_{2}}_{2}\big]\,,
\ee
and therefore cyclic permutations of the arguments vanish. We can also set $T^{A_1}=\mathbb{1}$, in which case we find another constraint:
\begin{align}
0\,&=\,
\mathcal{F}^{g^{h}}_\eta\big[ g^{h_{1}}_{1} g^{h_{2}}_{2} \dots g^{h_{n}}_{n}\big]
+
\mathcal{F}^{g^{h}}_\eta\big[ g^{h_{2}}_{2}g^{h_{1}}_{1} \dots g^{h_{n}}_{n} \big]
+
\dots\notag\\[2pt]
&\hspace{14pt}+
\mathcal{F}^{g^{h}}_\eta\big[  g^{h_{2}}_{2} \dots g^{h_{1}}_{1} g^{h_{n}}_{n} \big]
+
\mathcal{F}^{g^{h}}_\eta\big[  g^{h_{2}}_{2} \dots  g^{h_{n}}_{n} g^{h_{1}}_{1}\big]\,.
\end{align}
For $n=2$ and $n=3$, these $U(1)$ identities imply that there is only one distinct color-ordered form factor. 
We provide an explicit verification of these relations for $n=3$ in \cref{app:gggsplittingfunction}.
On top of the $U(1)$ decoupling, scattering amplitudes obey Kleiss-Kuijf relations  \cite{Kleiss:1988ne,DelDuca:1999rs} that further reduce the basis of independent structures and therefore imply additional relations for the form factors.


\subsection{MHV and N$^{n-1}$MHV Form Factors in QCD}
We now explain the MHV classification of helicity form factors for QCD.  As above, the MHV (and $\overline{\text{MHV}}$) form factors can be derived from attaching a single on-shell amplitude to the appropriate wavefunction factor, and are (anti-)holomorphic functions of the spinors.

\subsubsection{Scalar Form Factors}
We begin with the form factor for the scalar operator and a state consisting of a single scalar and $n-1$ gluons.
The one and two-point form factor have exactly the same form as the QED case in \cref{eq:FFscalar12} since neither the color ordering or the non-Abelian structure enters.
The first differences appear at three points.
We can compute the MHV form factor $\mathcal{F}^\phi_\eta[\phi_1 g_2^+ g_3^+]$ recursively either by performing a $\ra{\hat{3}}$-shift or a $\ra{\hat{2}}$-shift. 
For the $\ra{\hat{3}}$-shift, color ordering forces the shifted gluon to be adjacent to the operator insertion, and therefore there is only one possible on-shell diagram,

\begin{align}
		\includegraphics[width=0.35\linewidth]{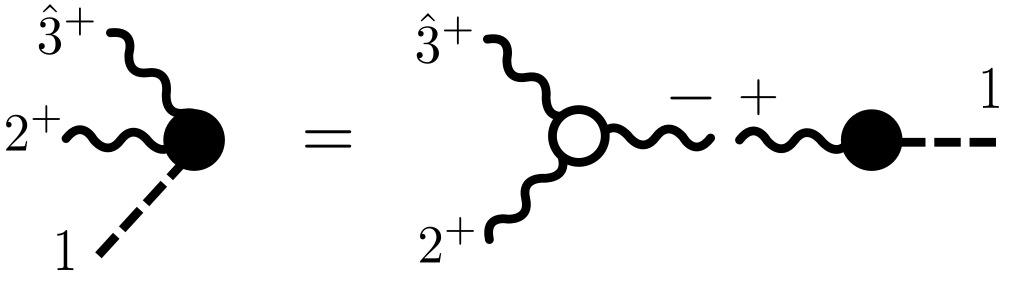}
\end{align}
In the case of the $\ra{\hat{2}}$-shift, one gets instead contributions from two on-shell diagrams, namely
\begin{align}
	\includegraphics[width=0.35\linewidth, valign=c]{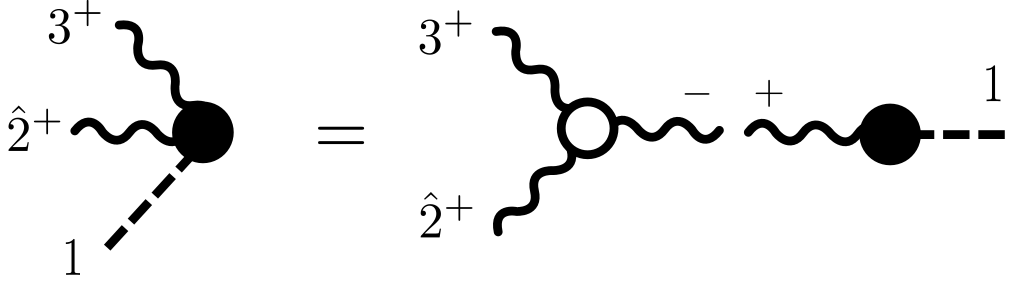}
\,\,	\,\,+\,
	\includegraphics[width=0.2\linewidth, valign=c]{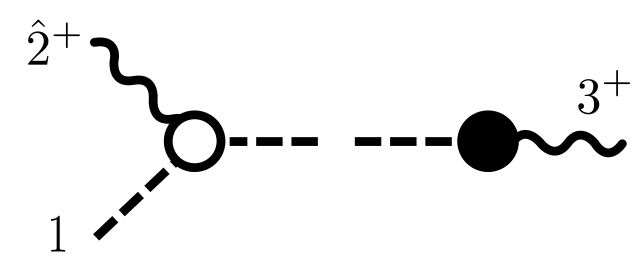}
\end{align}

Clearly, the $\ra{\hat{3}}$-shift is simpler. Nonetheless, both ways of constructing the form factor are of course equivalent and lead to
\be
\mathcal{F}^\phi_\eta[\,\phi_1\, g_2^+ g_3^+]
\,=\,
\frac{\ab{1 \eta }}{\ab{12}\ab{23}\ab{3\eta}}\,.
\ee

Similar to the QED case, the generic color ordered MHV form factor can be computed recursively and it is given by
\be
\includegraphics[width=0.09\linewidth, valign=c]{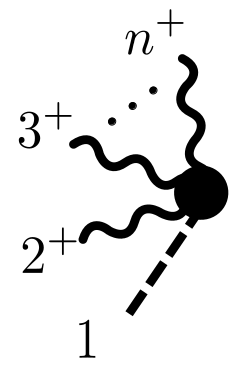}
\,=\,
\frac{\ab{1 \eta }}{\ab{12}\ab{23}\cdots \ab{n \eta}} \,.
\ee
Notice that the scalar QCD form factor only contains poles for the particles that are adjacent in the color ordering, while the QED one in \cref{eq:QEDscalarMHV} contains divergences when any photon is collinear to either the scalar or the Wilson line.
However, it is still true that the eikonal limit is trivial for the MHV form factor.  Following the same logic that led to \cref{eq:FFeikonalMHV} for the QED case above, we find
\be
\mathcal{F}_{\eta\rho}[g_2^+\dots g_n^+]
=
\lim_{\omega\to\infty} \mathcal{F}_{\eta}^\phi[\phi_1\,g_2^+\dots g_n^+]\vert_{p_1^\mu=\omega \rho^\mu}
=
\frac{\ab{\rho\eta}}{\ab{\rho 2}\ab{2 3}\dots \ab{n \eta}}\,.
\label{eq:FFeikonalMHVQCD}
\ee
Therefore, as in QED (see \cref{sec:SoftExchanges} above), the MHV form factor of a scalar field is identical to the one with a Wilson line in the direction of its momentum, indicating that the MHV form factor contains only soft radiation.  Note that the same form was found when computing a related form factor in $\mathcal{N}=4$ Super Yang Mills~\cite{Brandhuber:2010ad}.

\subsubsection{Fermion Form Factors}
Form factors with fermion operators and states containing a single fermion and an arbitrary number of gluons can be computed analogously. The MHV form factor is given by
\be
\mathcal{F}^{f^-}_{\eta} [f_1^- g_2^+\dots g_n^+]
=
\frac{\ab{1\eta}^2}{\ab{12}\ab{23}\cdots \ab{n \eta}}\,,
\ee
and the N$^{n-1}$MHV form factor is given by
\be
\mathcal{F}^{f^-}_{\eta} [f^- g^-\dots g^-]
=
\frac{\ls{\eta}q\ra{\eta}}{\sq{12}\sq{23}\cdots \sq{n \eta}}\,.
\ee
Again, note that both have a simple relation to the MHV and $\overline{\text{MHV}}$ scalar form factors:
\begin{subequations}
	\bea
	\mathcal{F}^{f^-}_{\eta} [f_1^- g_2^+\dots g_n^+]
&=&
\mathcal{F}^{\phi}_{\eta} [\phi_1 g_2^+\dots g_n^+]\, \ab{1\eta}\,,\\[3pt]
	\mathcal{F}^{f^-}_{\eta} [f_1^- g_2^-\dots g_n^-]
&=&\mathcal{F}^{\phi}_{\eta} [\phi_1 g_2^-\dots g_n^-]\, \frac{\ls{\eta}q\ra{\eta}}{\sq{1\eta}}\,,
	\eea
\end{subequations}
which coincides exactly with the same relations we found for the Abelian theory, see Eqs.~(\ref{eq:fermiontoscalarFFQED}).

\subsubsection{Gauge Boson Form Factors}
\label{sec:GBFormFactorsQCD}
In the case where the operator corresponds to a gluon, we can similarly compute the MHV form factor recursively. Starting from the operator as defined in \cref{eq:gluonFFdef}, which leads to the one point form factors in \cref{eq:gluonFFdef1point}, the MHV form factor is
\be
\mathcal{F}^{g^-}_\eta[g_1^+\dots g_i^-\dots g_n^+]
\,=\,
\frac{\ab{i\eta}^4}{\ab{\eta 1}\ab{12}\dots \ab{n\eta}}\frac{1}{\ls{\eta}\slashed q\ra{\eta}}\,.
\label{eq:MHVgluonformfactor}
\ee
Notice the non-holomorphic factor $1/\ls{\eta}\slashed q\ra{\eta}$.
This is expected, since the form of the operator leads to a one-point form factor that is already  non-holomorphic, see \cref{eq:gluonFFdef1point}.\footnote{
We note the following curiosity.  Choosing the operator as in \cref{eq:gluonFFdef} is customary in the modern literature of soft-collinear factorization, see \emph{e.g.}~\cite{Becher:2014oda, Hill:2002vw, Agarwal:2021ais}. 
Notice however, that an operator like $G_{\mu\nu}\eta^\mu \eta_\perp^\nu$ accompanied by Wilson lines, is similar to what was considered in earlier papers \cite{Collins:1981uw,Collins:1989gx}, and leads to the one point form factor given by 
\be
\la{g^-} W_\eta^\dagger\, G_{\mu\nu}\eta^\mu \eta_\perp^\nu\ra{0}
\,=\,
\ab{1\eta}^2\,,
\ee
which is holomorphic. Using this as a seed, the MHV form factor obtained is given by
\be
\la{g_1^+\dots g_i^-\dots g_n^+} (W_\eta^\dagger\, G_{\mu\nu}\eta^\mu \eta_\perp^\nu)
\ra{0}
\,=\,
\frac{\ab{i\eta}^4}{\ab{\eta 1}\ab{12}\dots \ab{n\eta}}\,,
\ee
which is holomorphic.}

Following the convergence arguments in \cref{sec:Recursion}, one would conclude that all form factors generated by $\mathcal{O}_g$ can be constructed using one line shifts except the one where the helicity of the gluon generated by the operator is the same as all the radiated gluons $\mathcal{F}^{g^-}_\eta[g^-\dots g^-\dots g^-]$,  the N$^{n-1}$MHV form factor.
Specifically, the issue is that the arguments for the constructibility of the N$^{n-1}$MHV form factor fail for a single line shift because this form factor does not have any external gluons with the opposite helicity from the operator. It therefore seems that one would need to use two-line shifts to compute these form factors recursively. 
Surprisingly, we will now show that the N$^{n-1}$MHV form factor actually converges faster than what the argument above suggests, demonstrating that a single line shift can be used to construct it recursively.

We can construct the two-point form factor $\mathcal{F}^{g^-}_\eta[g^-_1 g^-_2]$ using the two line shift with $\rs{\hat{1}}=\rs{1}+z\rs{\eta}$ and $\rs{\hat{2}}=\rs{2}+z\rs{\eta}$. This gives
\be
\includegraphics[width=0.25\linewidth, valign=c]{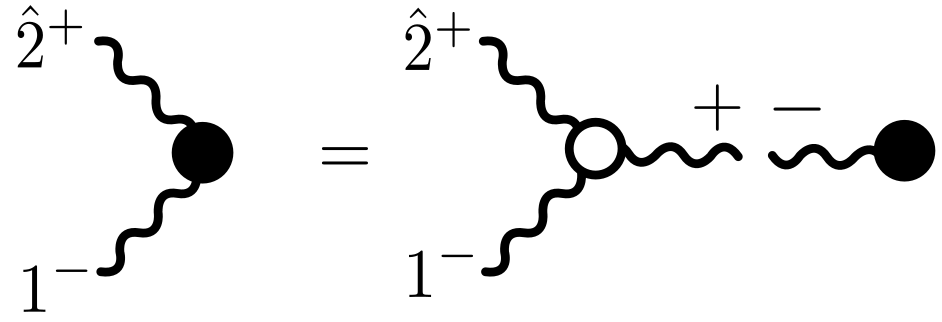}
\,\,\,=\,\,\,
\frac{\la{\eta}\slashed q\rs{\eta}}{\sq{\eta 1}\sq{12}\sq{2\eta}}\,,
\ee
where $q^\mu = p_1^\mu + p_2^\mu$. Applying either a $\rs{\hat{1}}$ or a $\rs{\hat{2}}$ single line shift to this result, we see that this form factor goes to zero as $z^{-1}$ for $z\to \infty$.  The reason is that 
the numerator is independent of $z$, so it is better behaved than the naive argument above would suggest.  
Since the numerator is simply a function of $\rs{\eta}$ there is no $z$-dependence under any choice of one-line shift.  This can already be seen in the one point form factor in \cref{eq:gluonFFdef1point}, where the numerator does not introduce any $z$ dependence.

We can therefore compute the N$^{n-1}$MHV gluon form factor with any number of external legs using a single-line shift or a two-line shift.\footnote{Shifting $\ra{\hat{1}}$ and $\rs{\hat{n}}$ makes two-line shift calculations simpler, but not as simple as the single-line shift.}  We find the same expression in both cases:
\be
\mathcal{F}^{g^-}_\eta[g^-\dots g^-]
\,=\,
\frac{q\cdot \eta}{\sq{\eta 1}\sq{12}\dots \sq{n\eta}}\,,
\label{eq:Nnm1MHVformfactorgluon}
\ee
where the numerator is the result of contracting $\slashed q\rs{\eta}\ls{\eta}$ with the projector $P_{\eta^+}$. 

As above, both forms lead to a simple relation between the MHV and N$^{n-1}$MHV gluon form factors and the scalar MHV and $\overline{\text{MHV}}$ ones:
\begin{subequations}
\bea
\mathcal{F}^{g^-}_{\eta} [g^- g^+\dots g^+]
&=&
\mathcal{F}^{\phi}_{\eta} [\phi g^+\dots g^+]\, \frac{\ab{1\eta}^2}{q\cdot\eta}\,,\\[5pt]
\mathcal{F}^{g^-}_{\eta} [g^- g^- \dots g^-]
&=&\mathcal{F}^{\phi}_{\eta} [\phi g^-\dots g^-]\, \frac{q\cdot\eta}{\sq{1\eta}^2}\,.
\eea
\end{subequations}

\subsection{N$^k$MHV Structure}
\label{sec:NkMHV}

Now we investigate the structure of the helicity form factors at the N$^k$MHV level for generic $k$. We focus on the form factor associated with the operator interpolating a negative helicity gluon and a state containing $n$ gluons where $k+1$ of them have negative helicity:
\be
\mathcal{F}^{\text{N}^k\text{MHV}(n)}_\eta\equiv \mathcal{F}^{g^-}_{\eta} [\,\underbrace{g^-  \dots g^-}_{k+1}\underbrace{\,g^+  \dots g^+}_{n-k-1}\,]\,.
\ee
The arguments below does not depend on the gluon ordering.
The recursion relation induced by the single-line shift allows to write this form factor in terms of an amplitude with $n_\mathcal{A}$ gluons, $k_\mathcal{A}$ of them with minus helicity, and a form factor with $n_{\mathcal{F}}$ gluons and $k_\mathcal{F}-1$ of them with negative helicity. The structure of the recursion relation implies that $n=n_\mathcal{A}+n_\mathcal{F}-2$ and $k=k_\mathcal{A}+k_\mathcal{F}-1$, since the internal propagator absorbs two legs, and one must necessarily be associated with a negative helicity gluon.

Consider the MHV form factor, and therefore $k=0$ and $k_\mathcal{A}+k_\mathcal{F}=1$. Since the MHV structure of the amplitudes requires $k_\mathcal{A}\geq 1$, we necessarily have $k_\mathcal{F}=0$ and indeed the only required form factor is the MHV one. Moreover, $k_\mathcal{A} = 1$ is only allowed to be non-zero due to the special kinematics of three point amplitudes with complex momentum.  We therefore have $n_\mathcal{A}=3$ and thus $n=n_\mathcal{F}+1$, which is indeed the result used to compute the MHV form factors.

At the NMHV level, $k=1$ and the two solutions $(k_\mathcal{F}, k_\mathcal{A})$ are either $(0,2)$ or $(1,1)$. In the first case, the decomposition is in terms of an MHV form factor and an MHV amplitude. In the second, an NMHV form factor is required, but again forcing a three point amplitude and therefore a lower point NMHV form factor.
At level $k$, there is a sum of contributions, starting from an N$^k$MHV amplitude with an MHV form factor, and ending on a 3-point amplitude and an $n-1$ point N$^k$MHV form factor.

For $k$ close to $n-1$, the recursion relation gets simpler again, as we have seen above. Indeed, for $k=n-1$, the requirement is $k_\mathcal{A}+k_\mathcal{F}=n_\mathcal{A}+n_\mathcal{F}-2$. This can only be satisfied for $k_\mathcal{F}=n_\mathcal{F}-1$ and $k_\mathcal{A}=n_\mathcal{A}-1$, which corresponds to an N$^{n-1}$MHV form factor and a 3-point amplitude, again as shown in the derivation of the explicit formula above.  For $k=n-2$, the two contributions come from N$^{n-1}$MHV form factors with $\overline{\text{MHV}}$ amplitudes and an N$^{n-2}$MHV form factor with a 3-point amplitude. This leads to a CSW-like expansion where the MHV and N$^{n-1}$MHV form factors are determined in a very simple manner from their lower points and the three point amplitudes, and act as building blocks in order to construct form factors that are their neighbor in $k$.
\clearpage

\subsection{All-line Shift Constructability}
For completeness, we will now argue that an all-line shift can be used to construct the helicity form factors.  Note that in comparison with the all-line shift recursion relations for amplitudes studied in~\cite{Cohen:2010mi}, these form factors will depend on the ``reference spinor,'' since in our case here that spinor corresponds to the null vector that defines the physical Wilson line direction.

Specifically, we can show that N$^k$MHV form factors decay as $z^{-k}$ at large $z$. The single-line shift allows us to write the form factor as
\be
\mathcal{F}^{\text{N}^{k}\text{MHV}(n)}_\eta
\,=\,\sum
\hat{\mathcal{A}}^{\text{N}^{k_{\mathcal{A}}}\text{MHV}(n_\mathcal{A})} \times \frac{1}{P^2}\times
\hat{\mathcal{F}}^{\text{N}^{k_\mathcal{F}}\text{MHV}(n_\mathcal{F})}_\eta\,,
\ee
where the sum is over different channels is implicit and the ``hat'' indicates dependence on the complex parameter controlling the deformation from the single line shift. 
Using this representation of the amplitude, we can now perform an all-line shift
\be
\rs{\tilde{i}}\,=\,\rs{i} + z\s c_i\s \rs{\eta}\,.
\ee
The constants $c_i$ are such that momentum is preserved under the shift, so $\sum_i c_i \ra{i}\ls{\eta}=0$ when the sum runs over all momenta, but nonvanishing whenever the sum runs over any other subset of momenta.
Under this shift, the N$^k$MHV amplitude scales as $z^{-k}$ at large $k$, while the propagator $1/P^2$ scales as $z^{-1}$ \cite{Elvang:2008vz}. 
Using the fact that the MHV form factor is a holomorphic function of the angle brackets, it scales as $z^0$. 
Since $k=k_\mathcal{A}+k_\mathcal{F}+1$ (taking into account that the 3-point amplitude counts as $k=-1$), one can see by induction that the N$^k$MHV form factor indeed behaves as $z^{-k}$ at large $z$. One can verify this explicitly taking the form in \cref{eq:Nnm1MHVformfactorgluon}, which indeed scales as $z^{n-1}$.
This behavior implies the existence of bonus relations among form factors \cite{Spradlin:2008bu}.

\section{Outlook}
\label{sec:Conclusions}
In this work, we studied the form factors associated with collinear Wilson line dressed operators, which are ubiquitous in the factorization of the soft and collinear modes in gauge theories.
We presented a novel recursive approach to computing these important objects.  Our methods lead to a powerful computational tool allowing us to build the $n$-point form factor starting from the free theory wavefunctions.
The recursion relation presented in \cref{sec:recrelationFFs} 
provides an efficient strategy based on deriving the $n$-point form factors using the $n-1$-point form factors and on-shell amplitudes as input building blocks.  This is accomplished by relying on a single-shift of an external momentum into the complex plane in a specific way that avoids the poles from eikonal factors. This provides a new way to compute such objects with respect traditional methods, and leads to novel representations of the helicity form factors.
We presented examples for both QED in \cref{sec:QED} and for QCD in \cref{sec:QCD}. The non-vanishing form factors with the maximal number of same-helicity gluons are denoted MHV and N$^{n-1}$MHV form factors, and they have a very simple form, reminiscent of the simplicity of MHV and $\overline{\text{MHV}}$ amplitudes. 

The new perspective that this work introduces opens up many directions for future exploration.
The most obvious direction is to 
push the calculation to higher points, perhaps by automating the recursion relations. This would provide a cross check of the state of the art $1 \to 4$ calculations \cite{DelDuca:2019ggv,DelDuca:2020vst}, while also producing a new representation of the result that could both be more numerically efficient and also would provide the input to even higher point calculations via recursion.
In another direction, it would be very exciting to apply these methods to explore the (rapidity) renormalization group evolution of these form factors \cite{Becher:2010tm,Becher:2011xn,Chiu:2011qc,Becher:2011pf,Chiu:2012ir,Becher:2012qa}.  For example, one could attempt to extend the on-shell methods of \cite{Caron-Huot:2016cwu} in order to extract the rapidity anomalous dimensions of the helicity form factors.  The on-shell perspective on the renormalization group for amplitudes has been shown to make the non-renormalization structure of these calculations transparent, while also reducing the complexity of some multi-loop calculations \cite{EliasMiro:2020tdv}.  First steps have been accomplished recently in \cite{Rothstein:2023dgb}, and we are optimistic that extending these approaches would yield valuable insight into the nature of the Wilson line form factors.
Another interesting avenue would be to explore the consequences for energy correlators. 
On the one hand, the formalism presented provides the seed to compute the $n$-point energy correlator~\cite{Basham:1977iq, Basham:1978bw, Basham:1978zq, Basham:1979gh} of a quark or gluon jet. On the other hand, it would be interesting to compare the structure provided by the MHV classification with the underlying CFT structure of such correlators, both in QCD and $\mathcal{N}=4$ Super Yang Mills~\cite{Chen:2019bpb, Chen:2022jhb}.
Finally, we speculate that the techniques developed here could also be applied to study gravity, with relevance to previous explorations such as~\cite{Naculich:2011ry, White:2011yy, Beneke:2012xa, Okui:2017all, Chakraborty:2019tem, Parra-Martinez:2020dzs, Brandhuber:2023hhy, Kosower:2018adc, Cristofoli:2021vyo, Bastianelli:2021nbs, Bonocore:2021qxh,  Beneke:2021aip, Bonocore:2022blj, Beneke:2022ehj, Beneke:2022pue,  Elkhidir:2023dco}.

Understanding the properties of gauge theories from an on-shell perspective has led to remarkable insights into the structure of the amplitudes and led to a multitude of new insights into the nature of these theories.  Extending these approaches to form factors broadens the scope of these developments to a new class of pseudo-observables.  We are optimistic that this paper only begins to scratch the surface of what can be learned using these techniques. 

\acknowledgments

We would like to thank Samuel Abreu, Justin Berman, Henriette Elvang, Pier Monni, and Ian Moult for their invaluable feedback.
T.~Cohen is supported by the U.S.~Department of Energy under grant number DE-SC0011640. The package S@M has been used in this work \cite{Maitre:2007jq}.

\appendix
\addcontentsline{toc}{section}{\protect\numberline{}Appendix}%
\section*{Appendix}
\section{From Form Factors to Splitting Functions}
\label{sec:SplittingFuncs}

In this appendix, we show how to derive splitting functions using helicity form factors as inputs.
Take $\la{\alpha}=\la{p_1\dots p_n}$ in \cref{eq:FormFactorDef} to be the state with all momenta collinear to some direction $k_1^\mu$, $p_i\cdot k_1 \sim \theta^2$ with $\theta\ll 1$.  A scattering amplitude for this collinear configuration factorizes as 
\be
\mathcal{A}(p_1 \dots p_n, k_2 \dots k_m) = \sum_\mathcal{O} \la{p_1\dots p_n} W^\dagger_\eta\mathcal{O}\ra{0} \mathcal{A}(q, k_2 \dots k_m)\,+\, O(\theta^2)\,,
\ee
with $\sum_i p_i^\mu = q^\mu$, see \emph{e.g.}~\cite{Feige:2013zla,Freedman:2013vya}.
The sum runs over all operators that interpolate with the external state. At leading order in $\theta^2$ and in QCD, due to fermion number conservation there is no sum if the hard parton is a fermion, and a sum over helicities if it is a gluon.
From this expression, one can express the usual splitting functions $\hat{P}^{ss^\prime}_{1\dots n}$ as defined in \emph{e.g.}~\cite{Catani:1999ss} in terms of the form factors $\mathcal{F}_\eta^\mathcal{O}$:
\be
\left(2\frac{g^2}{ q^2}\right)^{n-1} \hat{P}^{ss^\prime}_{1\dots n} \,=\,
\mathcal{F}_\eta^{\mathcal{O}_s} (p_1\dots p_n)\big(\mathcal{F}_\eta^{\mathcal{O}_{s^\prime}} (p_1\dots p_n)\big)^\dagger\,,
\ee
where the initial parton splits into partons 1 to $n$ with momenta $p_i$ and $q^\mu=\sum_i p_i$.
The possible spin interference appears since the same final state might be interpolated by operators with different helicity, denoted by $\mathcal{O}_s$ and $\mathcal{O}_{s^\prime}$. The only non-trivial interference occurs for the gluon operators. Note also that the MHV sector does not interfere.

Let us discuss the kinematics relevant for the splitting functions. Following the discussion in \cref{sec:Recursion}, we assume the sector to be collinear to some null direction $\bar{\eta}^\mu$. The opposite lightcone direction is denoted by $\eta^\mu$, and both define a basis of spinors as in \cref{eq:etaslashed}. 
The energy fractions $x_i$ are defined as
\be
x_i \,=\, \frac{2p_i\cdot \eta}{4Q}\,=\, \frac{\ab{\eta i}\sq{i \eta}}{\la{\eta} q \rs{\eta}}\,,
\ee
where $Q$ is the total energy of the state,
and therefore $\la{\eta} q \rs{\eta}= 4Q(x_1+\dots x_n) = 4Q$.

The invariants $s_{ij}=(p_i+p_j)^2$ can also be written in terms of an overall scale $Q^2$ times dimensionless quantities:
\be
s_{ij} = \ab{ij}\sq{j i}\,=\, 4 Q^2 x_i x_j \frac{1-\cos\theta_{ij}}{2} \,\equiv\, 4 Q^2 x_i x_j z_{ij}\,,
\ee
where $z_{ij}$ is zero in the strict collinear limit. Poles in $z_{ij}$ and in $x_i$ become collinear and soft singularities, respectively.
Note that the spinor products $\ab{i\eta}$ and $\sq{i\eta}$ are little group covariant and therefore carry important information about the helicity of the particle, crucial when considering processes at the amplitude level or observables sensitive to the helicity structure. 

\subsection*{$1\to 2$ Splitting Functions}

In the following, we derive the Altarelli-Parisi splitting functions \cite{Altarelli:1977zs}
from the form factors, both as a cross check and also to fix notation and concepts.
The simplest object we can consider is the form factor of a vector current. The leading order contribution for a state containing a quark pair and a gluon is given by
\be
\la{f_i^-\bar{f}_j^+g_A^-} \bar{\psi}\gamma^\mu \psi \ra{0}
\,=\,
-g\sqrt{2}T_{ij}^A \left( \frac{[12]}{[31][32]}\langle 1|\gamma_\mu |2] + \frac{1}{[31]} \langle 3|\gamma_\mu |2]  \right)\,.
\label{eq:currentFFfull}
\ee
This expression is given by the sum of two Feynman diagrams, corresponding to radiating a gluon from each fermion leg.
The first term of the expression contains a collinear pole whenever the gluon is collinear to either quark, while the second term only has a pole when the gluon is collinear to the quark with the same helicity. Also, while the first term has a soft singularity for $p_3\to 0$, the second term is finite.
Let us assume that the negative helicity quark and gluon are collinear to some direction $\bar{\eta}^\mu$. The leading behavior of the form factor is given by
\be
\hspace{-8pt}\la{f_i^-\bar{f}_j^+g_A^-} \bar{\psi}\gamma^\mu \psi \ra{0}
\,\simeq \,\mathcal{F}^{f^-}_{\eta}(1^- 3^-) \la{f_i^- \bar{f}_j^+} \bar{\psi}\gamma^\mu \psi \ra{0}
\,=\, 
g\sqrt{2}T_{ij}^A \frac{\la{\eta} 1+3\rs{\eta}}{\sq{31}\sq{3\eta}}\,\la{\bar{\eta}} \gamma^\mu \rs{2 }\,.
\ee
The hard process is replaced by $\la{\bar{\eta}} \gamma^\mu \rs{2 }$.  This has the interpretation that the non-collinear sectors cannot resolve the individual momenta that are collinear to the hard particle. Similarly, the Wilson line direction in the form factor is given by any non-collinear momenta, in this case $p_2$. Note that the form factor has a spinor index like a wavefunction, but is not contracted with the amplitude since there is a $\ra{\eta}\la{\bar{\eta}}$ projector in between, as in \cref{eq:FermionProjectors}. Therefore, one can use $\la{\bar{\eta}}$ as an ``external wavefunction'' for the hard process.

The splitting functions for a quark splitting into a quark and a gluon 
are recovered by squaring the form factors in \cref{eq:2pointfermionFFmp,eq:2pointfermionFFmm} with the appropriate normalization (an extra $1/Q$ factor in order to make the splitting function dimensionless)
\be
\includegraphics[width=0.25\linewidth, valign=c]{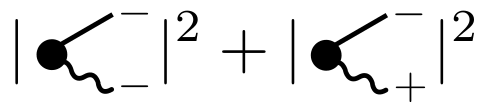}
\,\,\,=\,\,\,
C_F \left(  \frac{1}{x_2} + \frac{x_1^2}{x_2} \right)\,,
\ee
and the splitting functions $P_{qg}$ and $P_{gq}$ are obtained by writing $x_1=x,\, x_2=1-x$ and $x_1=1-x,\, x_2=x$, respectively.
The gluon splitting into gluons, is given by
\be
\includegraphics[width=0.4\linewidth, valign=c]{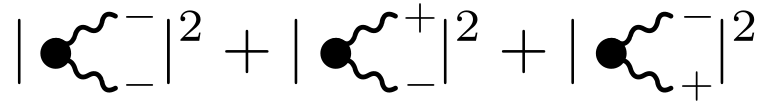}
\,\,\,=\,\,\,
2C_A \left(  \frac{1}{x(1-x)} + \frac{x^3}{1-x} + \frac{(1-x)^3}{x}  \right)\,.
\ee
Note that this splitting belongs to a minus helicity gluon form factor, but at this level the inclusive splitting function cannot discriminate between both helicities. 
Note also that the contributions coming from the MHV form factor contain singularities only when the plus helicity gluon goes to zero, while the N$^{n-1}$MHV form factor contains singularities at both $x\to 0$ and $x\to 1$.

The form factor of a gluon with a $q\bar{q}$ state is obtained by shifting any of the quarks, and given by
$\mathcal{F}^{g^-}_\eta(f^-_1\bar{f}^+_2) = \frac{\ab{1\eta}^2}{\ab{12}\la{\eta}q\rs{\eta}}$, which leads to
\be
\includegraphics[width=0.25\linewidth,valign=c]{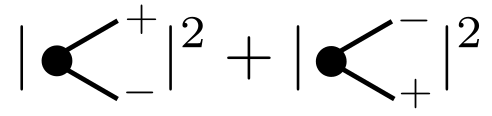}
\,\,\,=\,\,\,
T_R \left(  x + (1-x)^2 \right)\,.
\ee

\subsection*{$1\to 3$ Fermion Splitting Function}

We now reproduce the $1\to 3$ splitting function for a quark going to a quark and a quark-anti-quark pair of a different flavor. The form factor is given by
\be
\mathcal{F}^{f^-}_\eta(f_1^- f_2^{\prime -} \bar{f}_3^{\prime +})\,=\,\frac{1}{\la{\eta}q\rs{1}}\left(  \frac{\ab{2\eta}^2\la{\eta}q\rs{\eta}}{\ab{23}\la{\eta}2+3\rs{\eta}}
+
\frac{\la{\eta}q\rs{3}^2}{s_{123}\sq{23}}  \right)\,.
\ee
The two terms come from the product of the onshell quantities $\mathcal{A}[f^{\prime -} \bar{f}^{\prime +} g^\pm]\mathcal{F}^{f^-}_\eta[f^- g^\mp]$ and $\mathcal{A}[f^- \bar{f}^+ \bar{f}^{\prime +}]\mathcal{F}^{f^-}_\eta[f^-]$, respectively.
The square of the form factor
\begin{align}
	|\mathcal{F}|^2\,=\,
	\frac{1}{\text{tr}(\eta q 1 q)}\left( 
	\frac{s_{2\eta}^2s_{q\eta}^2}{s_{23}(s_{2\eta}+s_{3\eta})^2}
	+\frac{\text{tr}(\eta q 3 q)^2}{ s_{123}^2 s_{23}}
	+\frac{s_{q\eta }}{s_{123} s_{23}^2(s_{2\eta}+s_{3\eta})}
	(\text{tr}(\eta q 3 2)^2+\text{c.c.})
	\right)\,, \notag\\
\end{align}
leads to a representation for the splitting function equivalent to the usual one \cite{Catani:1998nv}. We used $s_{ab}=2p_a\cdot p_b$, with $q=p_1+p_2+p_3$, and
\be
\text{tr}(abcd)\,=\, 
\la{a}bcd\rs{a}\,=\, 
2( g^{\mu\nu}g^{\rho\sigma} - g^{\mu\rho}g^{\nu\sigma} + g^{\mu\sigma}g^{\rho\nu} + i \epsilon^{\mu\nu\rho\sigma})\,a_\mu b_\nu c_\rho d_\sigma\,.
\ee

\subsection*{$1\to 3$ Gluon Splitting Function}
\label{app:gggsplittingfunction}

As a final example, we compute the unpolarized $1\to 3$ splitting function for a gluon going to three gluons. The splitting function was originally derived in \cite{Campbell:1997hg,Catani:1999ss}. We first compute the different three-point form factors for the operator that corresponds to a negative helicity gluon, starting with the one point form factor
\be
\mathcal{F}^{g^-}_\eta[1^-] \,=\, \frac{\ab{1\eta}}{\sq{1\eta}}\,,
\ee
and the color ordered three point amplitudes
\be
\mathcal{A}[1^- 2^- 3^+]\,=\, \frac{\ab{12}^3}{\ab{23}\ab{31}}\,,
\qquad\text{and}\qquad
\mathcal{A}[1^+ 2^+ 3^-]\,=\, \frac{\sq{12}^3}{\sq{23}\sq{31}}\,.
\label{eq:threepointgluonamplitudes}
\ee
We are using an obvious shorthand $\mathcal{F}^{g^-}_\eta[1^-] = \mathcal{F}^{g^-}_\eta[g_1^-]$, etc.
The color ordered two-point NMHV form factor $\mathcal{F}^{g^-}_\eta[1^- 2^-]$ can be decomposed into lower points via either a $\rs{\hat{1}}$-shift or a $\rs{\hat{2}}$-shift. Either way, only one diagram contributes 
\be
\mathcal{F}^{g^-}_\eta[1^- 2^-] \,=\,
\mathcal{A}[1^- \hat{2}^- \hat{P}^+]\frac{1}{\ab{12}\sq{21}}\mathcal{F}^{g^-}_\eta[-\hat{P}^-]
\,=\,
-\frac{\langle \eta q \eta]}{\sq{\eta 1}\sq{12}\sq{2\eta}}\,,
\ee
where $q^\mu$ denotes the sum of all momenta in the form factor, $q^\mu=p_1^\mu + p_2^\mu$, and $\langle \eta q \eta] \equiv \la{\eta}(\slashed p_1+ \slashed p_2) \rs{\eta} = 2\eta\cdot q$.
The MHV form factor $\mathcal{F}^{g^-}_\eta[1^- 2^+]$ can be computed from a $\ra{\hat{2}}$-shift, and one gets
\be
\mathcal{F}^{g^-}_\eta[1^- 2^+] \,=\, -\frac{\ab{\eta 1}^3}{\ab{12}\ab{2\eta}}\frac{1}{\langle \eta q \eta]}\,.
\ee
The permutation $\mathcal{F}^{g^-}_\eta[1^+ 2^-]$ can also be computed via a $\ra{\hat{1}}$-shift, but it is unnecessary since the $U(1)$ decoupling forces the relation $\mathcal{F}^{g^-}_\eta[12]=-\mathcal{F}^{g^-}_\eta[21]$ for any choice of helicities, see Sec.~\ref{sec:QCD}.

Using the two point form factors and the four point amplitudes, one can construct the three-point form factors. The N$^{n-1}$MHV one can be computed by shifting either momenta and is given by
\be
\mathcal{F}^{g^-}_\eta[1^- 2^- 3^-] \,=\,
-\frac{\langle \eta q \eta]}{\sq{\eta 1}\sq{12}\sq{23}\sq{3\eta}}\,.
\ee
As explained in Sec.~\ref{sec:QCD}, the most efficient way to obtain the expression is by shifting either $\rs{\hat{1}}$ or $\rs{\hat{3}}$, since color ordering implies that one only diagram is nonvanishing. Similar reasoning applies to the MHV form factor. It is given by
\be
\mathcal{F}^{g^-}_\eta[1^- 2^+ 3^+] \,=\, -\frac{\ab{\eta 1}^4}{\ab{\eta 1}\ab{12}\ab{23}\ab{3\eta}}\frac{1}{\langle \eta q \eta]}\,.
\ee
The other form factors, $\mathcal{F}^{g^-}_\eta[1^+ 2^- 3^+]$ and $\mathcal{F}^{g^-}_\eta[1^+ 2^+ 3^-]$, are given by the same expression but with $\ab{\eta 2}^4$ and $\ab{\eta 3}^4$ in the numerator, respectively. One can check that they obey the $U(1)$ decoupling identity $\mathcal{F}^{g^-}_\eta[123]+\mathcal{F}^{g^-}_\eta[231]+\mathcal{F}^{g^-}_\eta[312]=0$ due to the Schouten identity.

For the NMHV form factor, there is a single independent color configuration, given by \emph{e.g.}~$\mathcal{F}^{g^-}_\eta[1^+ 2^- 3^-]$:
\be
\mathcal{F}^{g^-}_\eta[1^+ 2^- 3^-]\,=\,
\frac{1}{\langle \eta q \eta q 3]}\left(  \frac{\ab{\eta 2}\langle \eta q \eta 2\rangle^2}{\ab{12}\langle 1 \eta 3]\langle \eta (1+2)\eta]} +\frac{1}{s_{123}}\frac{\langle \eta q 1]^3}{\sq{12}\sq{23}} \right)\,.
\ee
One can check that the form factor $\mathcal{F}^{g^-}_\eta[1^- 2^- 3^+]$ obtained via recursion relations is equal to $\mathcal{F}^{g^-}_\eta[3^+ 2^- 1^-]$ as determined by the $U(1)$ decoupling relations. Similarly, $\mathcal{F}^{g^-}_\eta[1^- 2^+ 3^-]$ can be computed by shifting $\ra{\hat{2}}$,
\bea\nn
\mathcal{F}^{g^-}_\eta[1^- 2^+ 3^-]\,&=&\,
\frac{\langle \eta q \eta]\,\ab{3\eta}^3}{\ab{23}\langle \eta q 1]\langle 2\eta 1]\langle \eta (2+3)\eta]}
+
\frac{\langle \eta q \eta]\,\ab{\eta 1}^3}{\ab{12}\langle \eta q 3]\langle 2\eta 3]\langle \eta (1+2)\eta]}\\[4pt]
&&\,+\,\,
\frac{1}{s_{123}}\frac{\langle \eta q 2]^4}{\sq{12}\sq{23}\langle \eta q \eta] \langle \eta q 3]\langle \eta q 1]}\,.
\eea
A powerful cross check is provided again by the $U(1)$ decoupling, since this form factor is indeed not independent and $0=\mathcal{F}^{g^-}_\eta[1^- 2^+ 3^-]+\mathcal{F}^{g^-}_\eta[2^+ 3^- 1^-]+\mathcal{F}^{g^-}_\eta[ 3^- 1^- 2^+]$. We numerically verified that this is the case.

With this, the splitting function is given by the sum of the squares of these form factors:
\bea\nn
\langle \hat{P}_{g_1g_2g_3}\rangle &=& 4 C_A^2 s_{123}\bigg( \left|\mathcal{F}[1^-2^-3^-]\right|^2+\left|\mathcal{F}[1^+2^-3^-]\right|^2+\left|\mathcal{F}[1^-2^+3^-]\right|^2+\left|\mathcal{F}[1^-2^-3^+]\right|^2\\
&&+\left|\mathcal{F}[1^+2^+3^-]\right|^2+\left|\mathcal{F}[1^-2^+3^+]\right|^2+\left|\mathcal{F}[1^+2^-3^+]\right|^2
\,+\,\text{5 perm.} \bigg)\,.
\eea
We have verified numerically that this result agrees with the literature~\cite{Campbell:1997hg,Catani:1999ss}.

\end{spacing}

\begin{spacing}{1.09}
\addcontentsline{toc}{section}{\protect\numberline{}References}%
\bibliographystyle{JHEP}
\bibliography{Wilson_Line_Recursion}
\end{spacing}

\end{document}